\documentclass[iop]{emulateapj}

\usepackage{apjfonts}
\usepackage{times}
\usepackage{graphicx}
\usepackage{natbib}
\usepackage{color}

\newcommand{\OIII}{\mbox{[O\,\textsc{iii}}]}

\newcommand{\kms}{km s$^{-1}$}
\newcommand{\Ha}{H$\alpha$}
\newcommand{\Hb}{H$\beta$}

\shorttitle{Ionized Gas Kinematics along the Large-Scale Radio Jet.}
\shortauthors{Le et al.}

\begin{document}

\title{Ionized Gas Kinematics along the Large-Scale Radio Jets in Type 2 AGNs}
\author{Huynh Anh N. Le$^{1}$}
\author{Jong-Hak Woo$^{1}$\altaffilmark{$\dagger$}}
\author{Donghoon Son$^{1}$}
\author{Marios Karouzos$^{1,2}$}
\author{Aeree Chung$^{3}$}
\author{Taehyun Jung$^{4}$}
\author{Evangelia Tremou$^{5}$}

\affil{$^{1}$Astronomy Program, Department of Physics and Astronomy, Seoul National University, Seoul 151-742, Republic of Korea; woo@astro.snu.ac.kr}
\affil{$^{2}$Nature Astronomy, Springer Nature, 4 Crinan Street, N1 9XW London, United Kingdom}
\affil{$^{3}$Department of Astronomy, Yonsei University}
\affil{$^{4}$Korea Astronomy and Space Science Institute}
\affil{$^{5}$Department of Physics and Astronomy, Michigan State University, East Lansing, MI 48824, USA}

\altaffiltext{$\dagger$}{Author to whom any correspondence should be addressed: woo@astro.snu.ac.kr}

\begin{abstract}

To investigate the connection between radio activity and AGN outflows, we present a study of ionized gas kinematics based on \OIII\ $\lambda$5007 emission line along the large-scale radio jet for six radio AGNs. These AGNs are selected based on the radio activity (i.e., $\mathrm{L_{1.4GHz}}$ $\geqslant$ 10$^{39.8}$ erg~s$^{-1}$) as well as optical emission line properties as type 2 AGNs. Using the Red Channel Cross Dispersed Echellette Spectrograph at the Multiple Mirror Telescope, we investigate in detail the [O III] and stellar kinematics. We spatially resolve and probe the central AGN-photoionization sizes, which is important in understanding the structures and evolutions of galaxies. We find that the typical central AGN-photoionization radius of our targets are in range of 0.9$-$1.6 kpc, consistent with the size-luminosity relation of \OIII\ in the previous studies. We investigate the \OIII\ kinematics along the large-scale radio jets to test whether there is a link between gas outflows in the narrow-line region and extended radio jet emissions. Contrary to our expectation, we find no evidence that the gas outflows are directly connected to the large scale radio jets.
\end{abstract}


\section{Introduction}\label{intro}

The scaling relations between the mass of supermassive black holes and their host galaxy properties gave rise to the black hole-galaxy coevolution paradigm, advocating that these two components grow interactively through some form of self-regulation (e.g., \citealp{Ferrarese&Merritt00}.; \citealp{Gebhardt+00}; \citealp{Kormendy&Ho13}; \citealp{Woo+13}).
As large-scale cosmological simulations (e.g., the Millennium Simulation, \citealp{Springel+05}) showed how galaxies grow over cosmic time, it soon became apparent that certain regulation of star formation is needed in order to reconcile the observed mass function of galaxies in the local universe with the one derived from these numerical simulations (e.g., \citealp{Croton+06}, \citealp{Sijacki+07}).

Exploring feedback from active galactic nucleus (AGNs) become an important key to understand this self-regulation of black holes and their host galaxy evolution (\citealp{Croton+06}; \citealp{Dubois+13}) since AGNs with their enormous energy output may play a critical role in this self-regulation.
The process of star formation regulation or suppression through mechanisms related to the AGN (coined as AGN feedback), has been suggested as the answer to both aforementioned problems. AGN feedback is generally argued to manifest itself in two distinct modes \citep[see][]{Fabian12}. The radiative-mode feedback is attributed to high Eddington ratio AGNs, whose effects are driven by deposition of energy in the inter-stellar medium (ISM) through the AGN radiative pressure. On the other hand, radio-mode feedback is associated with radio-AGNs and relies on the deposition of mechanical energy from a radio jet. In particular, \citet{Croton+06} showed that radio-mode feedback can effectively prevent gas cooling, hence, halt the growth of the galaxy. Several observational studies have shown this mechanical energy deposition in action (e.g., \citealp{McNamara+05}). Furthermore, in some cases radio jets have been associated with ionized and molecular gas outflows (e.g., \citealp{Nesvadba+11}, \citealp{Morganti+13}), which would offer an alternative, if not complimentary, way of suppressing star formation through the depletion of a galaxy's gas reservoirs.

The ionized gas outflows may be observationally the most straightforward signatures to investigate whether and to what degree AGN can affect the ISM of their host galaxy. The velocity of \OIII\ $\lambda$5007 emission line is of particular interest for probing such outflows (e.g., \citealp{Zamanov+02}, \citealp{Komossa+08}, \citealp{Cresci+15}, \citealp{Cicone+16}, \citealp{Harrison+16}, \citealp{Yuan+16}, \citealp{Karouzos+16}, \citealp{Bae+17}, \citealp{Concas+17}, \citealp{Perna+17}). The spatially-integrated \OIII\ lines of AGNs have been observed to be blue or redshifted with respect to low-ionization lines (e.g., \citealp{Boroson05}), which is presumably due to the combined effect of outflows and dust extinction (\citealp{Crenshaw+10}, \citealp{Bae&Woo16}). Furthermore, the line profile of \OIII\ is often seen to be very asymmetric with a broad wing component, indicating the presence of high velocity gas.
\citet{Woo+16} recently performed a census of ionized gas outflows using a large sample of $\sim$39,000 type 2 AGN out to z$\sim$0.3, by investigating the velocity shift and velocity dispersion of the \OIII\ line \citep[see also,][]{Bae&Woo14, Woo&Bae17}. By analyzing the spatially-integrated SDSS spectra, they found that the majority of luminous AGNs shows kinematic signatures of outflows, indicating the prevalence of gas outflows.
It was also revealed that there is a strong contrast of the \OIII\ velocity dispersion (2nd moment) between AGNs and star-forming galaxies \citep{Woo+16}. While star-forming galaxies present a relatively narrow \OIII\ line, AGNs show a broad range of \OIII\ line widths (50$-$500 km s$^{-1}$) (see also, e.g., \citealp{Crenshaw&Kraemer00}; \citealp{Boroson05}; \citealp{Liu+13}; \citealp{Rodriguez+13}; \citealp{Karouzos+16}). This indicates the dramatic difference of ionized gas kinematics between AGNs and non-AGN galaxies due to non-gravitational kinematics, i.e., AGN-driven outflows.

The connection between radio activity and gas outflows is a topic of active research. While individual radio galaxies show
connection between jet and gas kinematics, gas outflows in most AGNs are not due to radio activity since a majority of AGNs are not
strong radio sources. Using radio and non-radio AGNs, \citet{Woo+16} reported that while the velocity and velocity dispersion of
ionized gas traced by \OIII\ dramatically increases with \OIII\ luminosity, which is an indicator of AGN activity, the kinematics of
ionized gas show no trend with radio luminosity, which is a tracer of jet activity, indicating that outflows are not driven by jet.
Nevertheless, AGNs with stronger outflows include more radio sources than AGNs with weaker outflows. For example, \citet{Bae&Woo14} found a $\sim10\%$ radio detection rate (at a 1.4 GHz flux density limit of 1 mJy from the FIRST radio survey) in their type 2 AGN sample, while this rate quadruples to $\sim40\%$ for those AGN with large \OIII\ velocity shift. This suggests an implicit connection between the ionized gas outflows and large-scale radio jets.

Among the SDSS AGNs used in our previous works (\citealp{Woo+16}; \citealp{Woo&Bae17}), the radio sources with the 1.4 GHz luminosity larger than 10$^{40}$ erg~s$^{-1}$ can be classified as radio-AGNs, which may host a radio-jet \citep{Mauch&Sadler07}. These relatively high luminosity radio sources with kinematic signature of ionized gas outflows are
a unique sample to study the connection between the ionized gas outflows and the large-scale radio jets.
Spatially resolved measurement of the narrow-line region (NLR) can provide a better understanding of the nature of AGN-driven outflows and their connection to radio activity by providing spatial and energetic links between these two components.

In this paper, we present the spatially resolved kinematic study of six radio AGNs, which are selected from our previous study of $\sim$39,000 optical type 2 AGNs at z $<$ 0.3 \citep{Woo+16}. By using the long slit data obtained from the Multiple Mirror Telescope (MMT), we investigate \OIII\ and stellar kinematics along the radio jet direction. In Section \ref{obs}, we describe our observations, sample selection, and data reduction processes. Section \ref{analy} presents our analysis methods. We discuss our results in Section \ref{result}.  Finally, Section \ref{sum} presents the summary and conclusions. Throughout the paper, we used the cosmological parameters as: $H_0 = 70$~km~s$^{-1}$~Mpc$^{-1}$, $\Omega_{\rm m} = 0.30$, and $\Omega_{\Lambda} = 0.70$.

\section{Observations}\label{obs}

\subsection{Sample Selection}

We used the type 2 AGN catalogue from \citet{Woo+16}, which contains a sample of $\sim$39,000 type 2 AGNs at z $<$ 0.3. These AGNs were classified based on the emission line flux ratios using the SDSS DR7 data, and \OIII\ kinematics were in detailed investigated to study AGN-driven outflows in comparison with star-forming galaxies \citep{Woo&Bae17}. To select radio AGNs for this study, we first cross-matched the type 2 AGN catalogue with the catalogue of the FIRST (Faint Images of the Radio Sky at Twenty Centimeters) survey at a 1.4 GHz, which has a flux density limit of 1 mJy \citep{Becker+95} \footnote  {http://sundog.stsci.edu/first/catalogs.html (14dec17 version)}. We found that $\sim$16$\%$ of optical type 2 AGNs (i.e., $\sim$6300 objects) show radio-luminosity at 1.4 GHz in range of 10$^{37}$$-$10$^{42}$ erg~s$^{-1}$. Second, among these radio-detected sources, we selected 360 targets, which have relatively high radio-luminosity ($\mathrm{L_{1.4GHz}}$ $\geqslant$ 10$^{39.8}$ erg~s$^{-1}$, see \citealp{Mauch&Sadler07}), extinction-corrected \OIII\ luminosity (L$_{\OIII}$ $>$ 10$^{40}$ erg~s$^{-1}$), and \OIII\ velocity dispersion ($\sigma_{\OIII}$ $>$ 130 km s$^{-1}$), in order to study the connection between large-scale radio jets and ionized gas outflows (see Figure 1). Third, for the selected 360 targets, we carefully examined FIRST images, which provide a 5\arcsec spatial resolution, to check the signatures of extended ($>$ 1$\arcsec$) radio-jet structures, and selected a sample of 16 most radio luminous targets (e.g., the peak flux density $>$ 33 mJy).

We were able to observe six objects out of the sample of 16 radio AGNs (see next Section). These AGNs have various levels of ionized gas outflows as the gas to stellar velocity dispersion ratio $\sigma_{\OIII}$/$\sigma_*$ ranges from 0.7 to 1.9. Note that the ratio larger than unity indicates that the kinematics of the ionized gas is influenced by the non-gravitational component such as outflows \citep{Woo+16} while the ratio $\leq$1 suggests that the gas kinematics is entirely governed by the gravitational potential of the host galaxy.  Among the six objects, two AGNs have relatively strong outflows, two AGNs have weak outflows, and the other two objects have no signature of gas outflows (see Table 1). Although the sample is composed of only six objects, the dynamic range of the ionized gas outflows may shed a light on the connection with radio emission.
Table \ref{tab:sample} provides the properties of the sample, which are adopted from \citet{Woo+16}. The \OIII\ luminosity is corrected for dust-extinction by using the extinction law in equation (1) of \citet{Calzetti99}. In this calculation, the ratio \Ha\ / \Hb\ is assumed to be 2.86 \citep{Netzer09}. More details can be found in \citet{Bae&Woo14} and \citet{Woo+16}. Note that for the AGNs with weak \Hb\ line, the correction factor for the dust extinction is highly uncertain since the measured \Hb\ flux has a large uncertainty due to the low S/N. Thus, we present both extinction-corrected and uncorrected luminosities for each object.

\subsection{Observations and Data Reduction}

We observed the sample on March 16th 2015, using the Red Channel Cross Dispersed Echellette Spectrograph at the Multiple Mirror Telescope (MMT) observatory. The echellette grating covers a wavelength range, 4300$-$8900 \AA\ with a spectral resolution of $\sim$90 km s$^{-1}$. Depending on the wavelength range, we obtained the wavelength scale 0.6$-$0.9 \AA\ pixel$^{-1}$. The slit-size was fixed at 1$\arcsec$ $\times$ 20$\arcsec$ with a spatial scale $\sim$0.3 $\arcsec$ pixel$^{-1}$. The seeing during the observation was $\sim$1.0$\arcsec$, and we observed each target at low airmass, e.g., $\sim$1.0-1.4.
%(see the log of observations in Table 1).

Based on the radio images, we determined the position angle (PA) of the slit along the jet to investigate whether the gas outflows are connected with the jet activity. For two objects, J104030$+$295758 and J130347$+$191617 (hereafter J1040 and J1303), we were not able to clearly define the jet direction, and used the morphology of the radio flux distribution
to decide the PA (see Figure \ref{fig:obs}). Using the 20\arcsec-long slit, we were able to probe the kinematics of the ionized gas at 10$-$20 kpc scales along the jet direction from the center of each galaxy. However, the low S/N in the obtained spectra from the outer part of these galaxies is the main limitation and we were able to investigate mainly the central part (see Section 4.3).

%%%%%%%%%%%%%%
\begin{figure}
\figurenum{1}
\centering
    \includegraphics[width=0.4\textwidth]{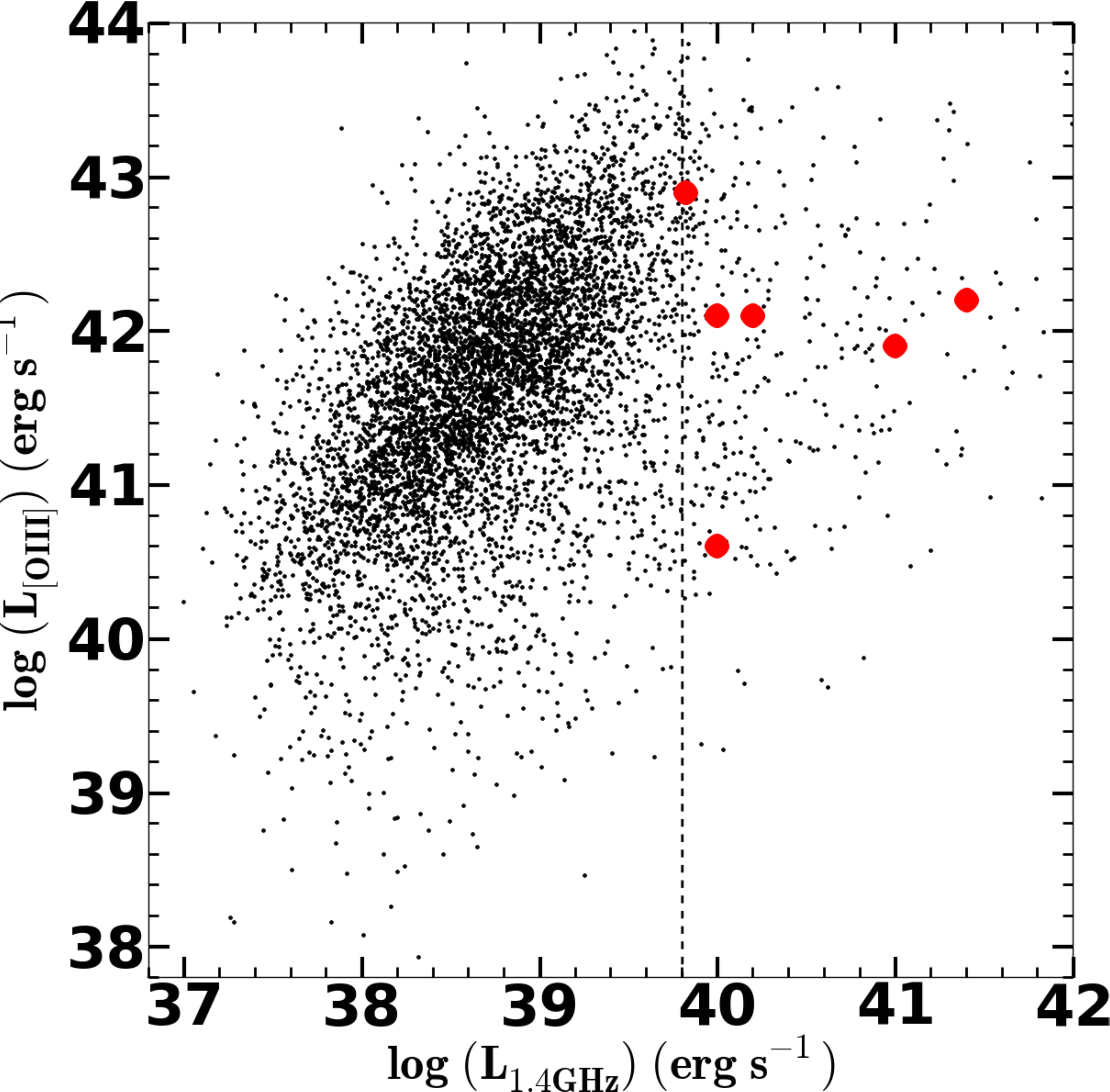}\\
    \caption{
    Sample selection targets for observations. The black-dots indicate $\sim$6,300 all radio-detected SDSS type 2 AGNs at z $<$ 0.3.
    The selected targets are shown in red-dots. The dash-line shows the radio-luminosity $\mathrm{L_{1.4GHz}}$ $\geqslant$ 10$^{39.8}$ erg~s$^{-1}$.
    \label{fig:sample}}
\end{figure}
%%%%%%%%%%%%%%%
%%%%%%%%%%%%
\begin{figure*}
\figurenum{2}
\centering
    \includegraphics[width=0.32\textwidth,height=0.32\textwidth]{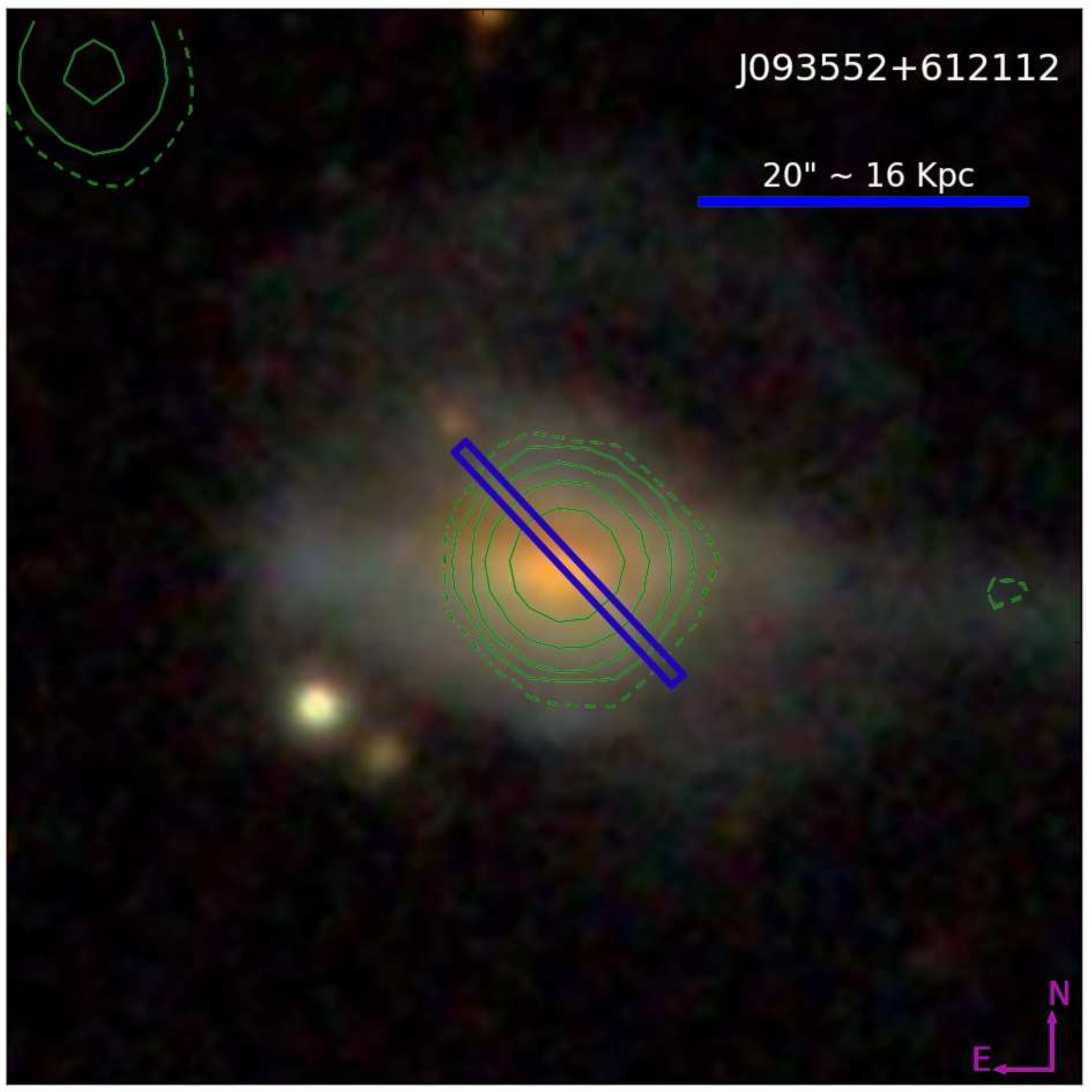}
    \includegraphics[width=0.32\textwidth,height=0.32\textwidth]{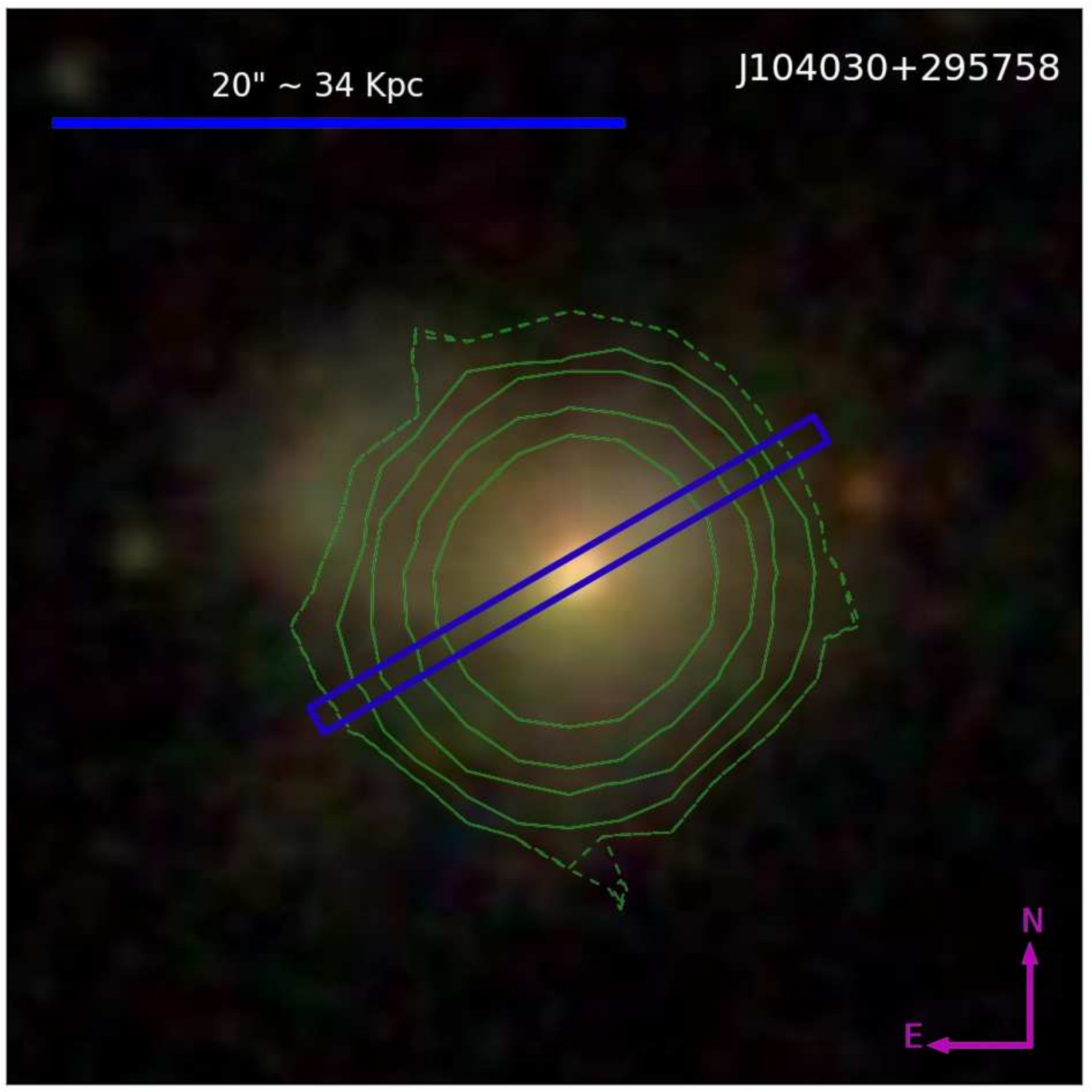}
    \includegraphics[width=0.32\textwidth,height=0.32\textwidth]{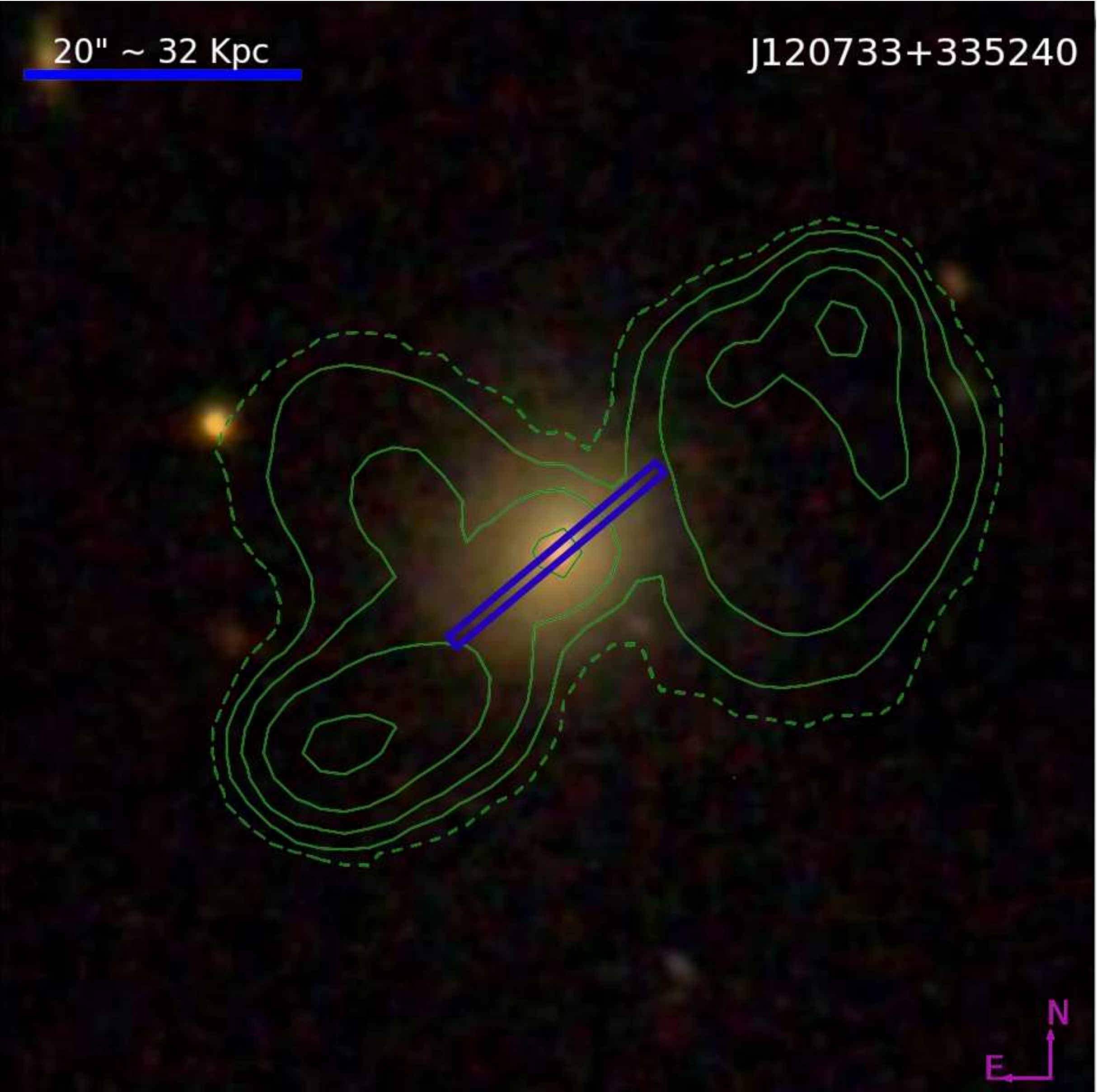}\\
    \includegraphics[width=0.32\textwidth,height=0.32\textwidth]{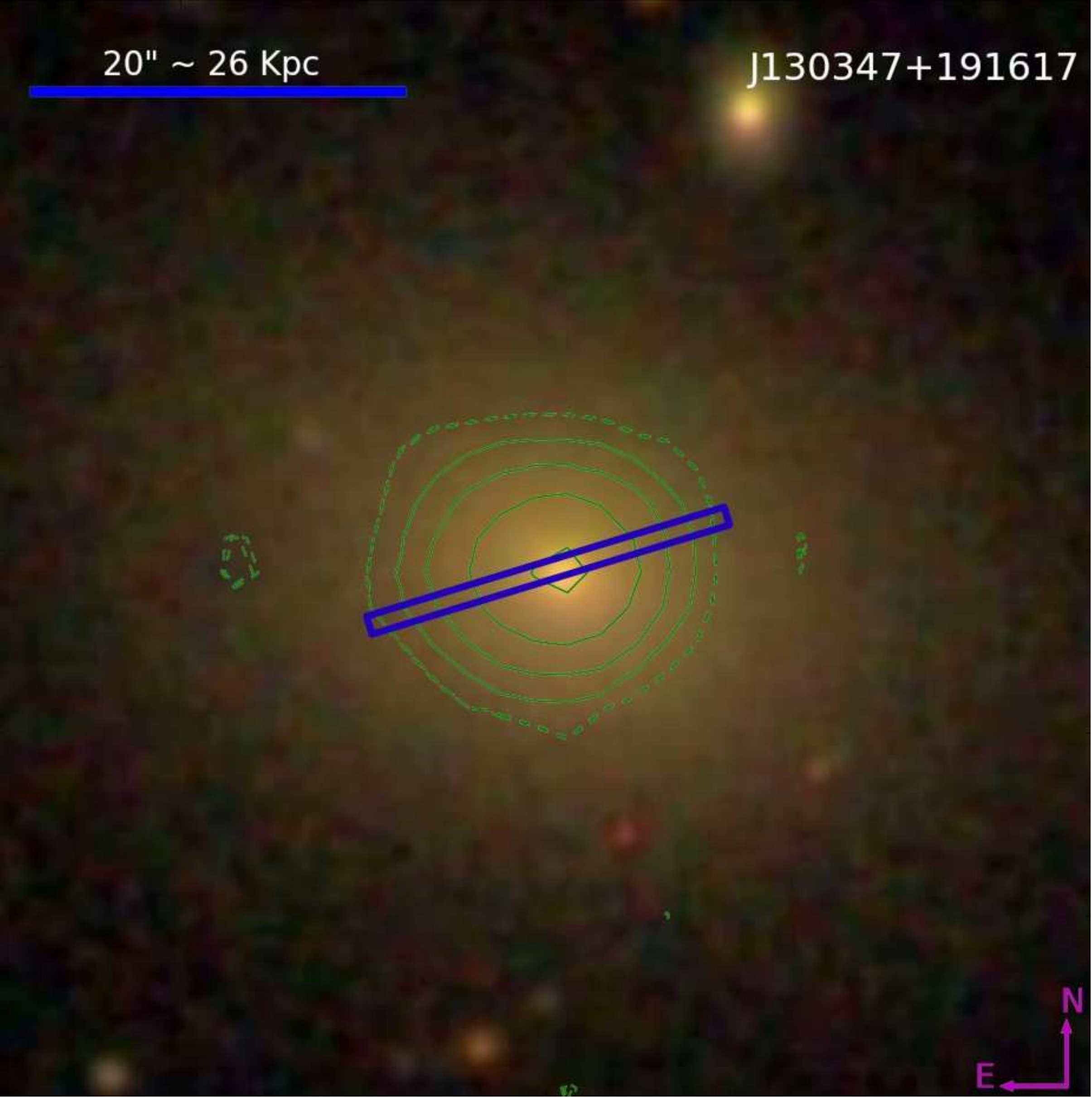}
    \includegraphics[width=0.32\textwidth,height=0.32\textwidth]{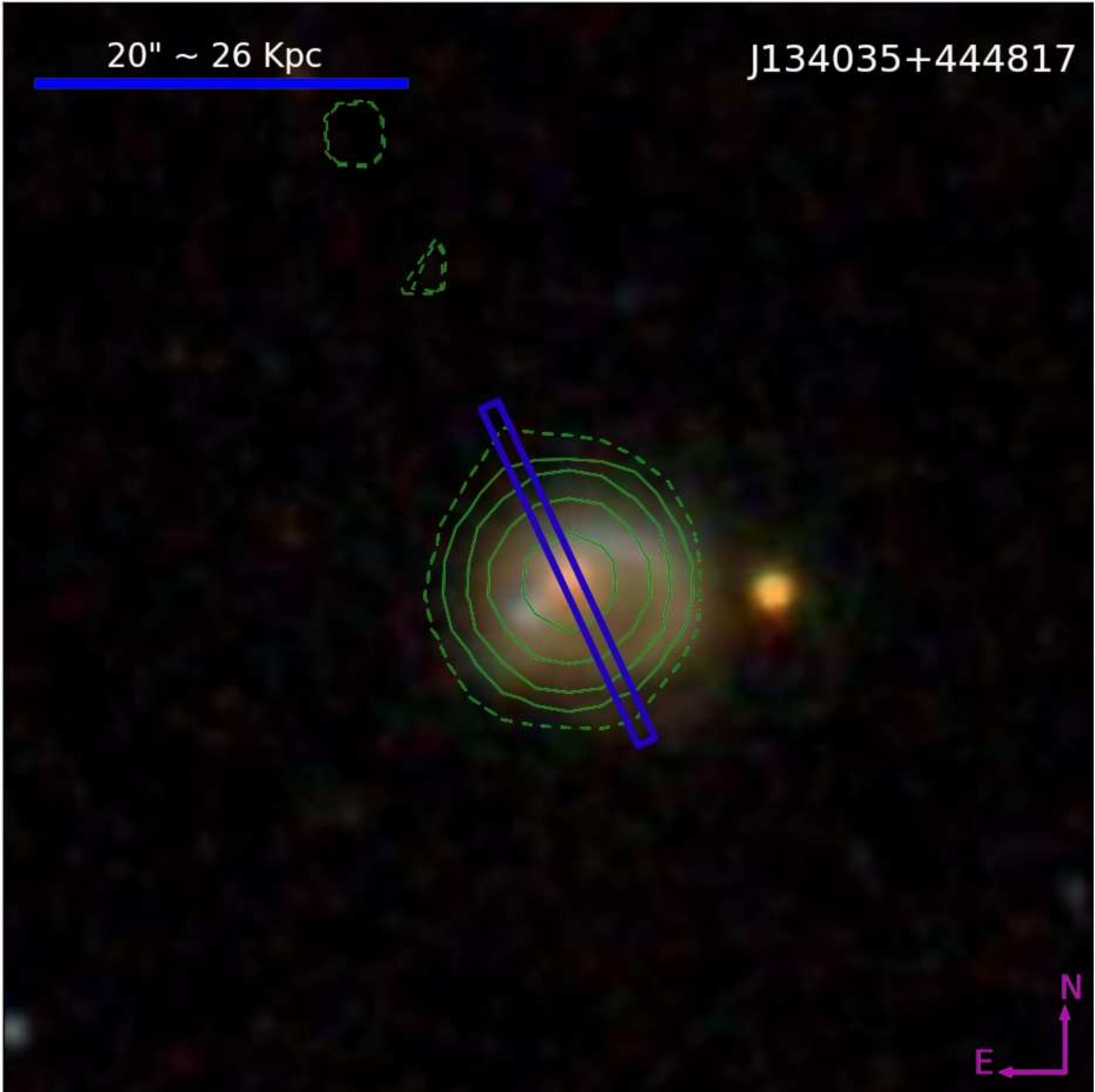}
    \includegraphics[width=0.32\textwidth,height=0.32\textwidth]{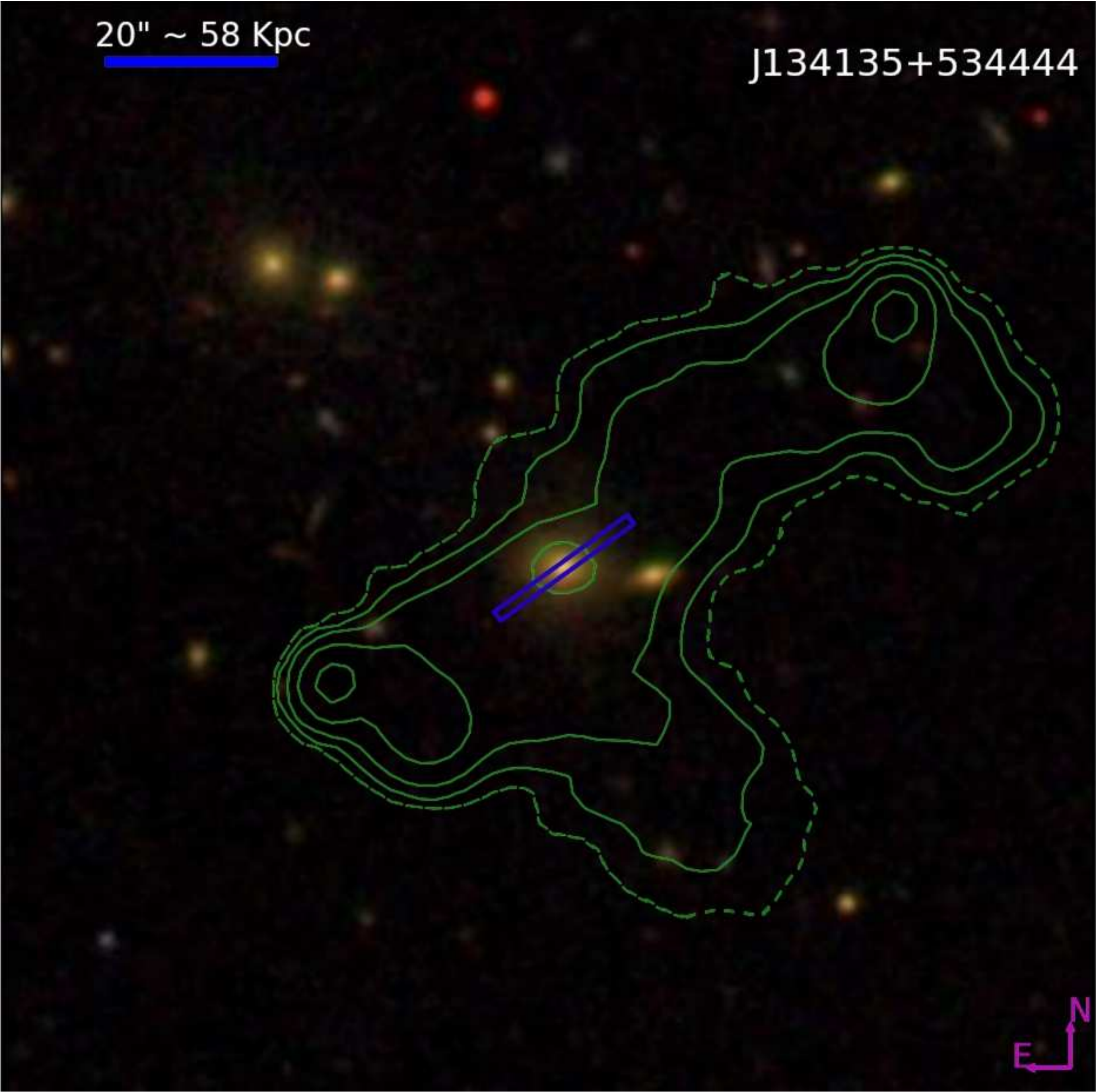}
    \caption{SDSS {\it gri} composite image of each target along with the slit position (1$\arcsec$ $\times$ 20$\arcsec$; thin blue box). 
    The 1.4 GHz density contours from the FIRST survey are presented with green solid lines (3$\sigma$ with dashed lines while 10, 30, 100, 300 $\sigma$ contours with solid lines). Note that we determined the jet direction based on the radio contours in a larger scale than presented here for four targets, while J1040 and J1340 show no clear jet direction.    
%The contour levels are shown based on n-sigma $\times$ the rms noise of each object (n-sigma in each object are, J093552$+$612112: 3, 30, 250, 442, 66 mJy; J104030$+$295758: 3, 33, 32, 467, 1105, 2105 mJy; J120733$+$335240: 3, 7, 20, 39, 66, 125, 171, 230, 309 mJy; J130347$+$191617: 3, 6, 25, 89, 190, 316 mJy; J134035$+$444817: 3, 14, 49, 98, 182, 336, 490 mJy; J134135$+$534444: 3, 5, 11, 27, 48, 64, 91, 128, 176, 246, 342 mJy) with the rms noise of 0.159 mJy, 0.152 mJy, 0.152 mJy, 0.158 mJy, 0.143 mJy, and 0.187 mJy, respectively. The green dash lines display the 3$\sigma$ rms noise of the radio 1.4 GHz density contours.
     \label{fig:obs}}
\end{figure*}
%%%%%%%%%%%%

%%%%%%%%%%%%%
\begin{table*}
\begin{center}
\tablewidth{\textwidth}
\tabletypesize{\scriptsize}
\caption{Properties of six type 2 AGNs} \label{tab:sample}
\begin{tabular}{ccccccccccccc}
\tableline\tableline
ID	&	$\alpha$ (J2000.0)	 &	$\delta$ (J2000.0)  &	z  &	 $\sigma$$_{\OIII}$ & ratio &  L$_{\OIII}$ &   L$_{\rm \OIII, unc}$ &  $\mathrm{L_{1.4GHz}}$  & t$_{\rm exp}$ & PA &  scale & seeing
\\
	&	(hh:mm:ss)       &  (dd:mm:ss)        &        	 & (km s$^{-1}$)  &  &  $\log$ (erg s$^{-1}$)  &  $\log$ (erg s$^{-1}$)  & $\log$ (erg s$^{-1}$) &  (min.) & (deg.) &  (kpc/$\arcsec$) &  ($\arcsec$) \\ 	
(1)   &   	(2)	       		   & (3)	                   & 	(4)                 &  (5)             	 &	(6)   &  (7) &   (8)  & (9)   & (10) & (11)  & (12) & (13) \\
\tableline
J093552$+$612112	&	09:35:52	&	$+$61:21:12	&	0.0392	&	211	&	1.12	&	42.6	& 40.0	& 39.8	&	55	&	 43	&	0.8	&	1.0	\\
J104030$+$295758	&	10:40:30	&	$+$29:57:58	&	0.0909	&	296	&	1.28	&	42.1	& 40.7	& 41.0	&	40	&	 -60	&	1.7	&	1.1	\\
J120733$+$335240	&	12:07:33	&	$+$33:52:40	&	0.0791	&	136	&	0.71	&	42.2	& 41.7	& 40.2	&	60	&	 -50	&	1.5	&	1.1	\\
J130347$+$191617	&	13:03:47	&	$+$19:16:17	&	0.0635	&	238	&	0.99	&	40.7	& 40.3	& 40.0	&	60	&	 -73	&	1.2	&	1.0	\\
J134035$+$444817	&	13:40:35	&	$+$44:48:17	&	0.0654	&	166	&	1.86	&	42.4	& 40.9	& 40.0	&	60	&	 24	&	1.3	&	1.1	\\
J134135$+$534444	&	13:41:35	&	$+$53:44:44	&	0.1410	&	413	&	1.72	&	42.1	& 41.3	& 41.4	&	90	&	 -54	&	2.5	&	1.0	\\

\tableline
\end{tabular}
\tablecomments{Column (1) Sample ID; Column (2) \& (3) Right ascension and declination; Column (4) Redshift of targets based on SDSS; Column (5) \OIII\  velocity dispersion measured from the SDSS spectra. Column (6) The ratio of \OIII\ velocity dispersion and stellar velocity dispersion ($\sigma$$_{\OIII}$$/$$\sigma$$_{*}$); Column (7) Dust-corrected \OIII\ luminosity; Column (8) Dust-uncorrected \OIII\ luminosity; Column (9) Radio luminosity at 1.4 GHz; Column (10) Total exposure time (minutes); Column (11) Slit position angles (degrees); Column (12) Linear scale along the slit direction; Column (13) Seeing during the observation.}
\end{center}
\end{table*}
%%%%%%%%%%%%%

We performed the standard data reduction process with the IRAF packages\footnote{IRAF (Image Reduction and Analysis Facility) is distributed by the National Optical Astronomy Observatories (NOAO).}. First, we corrected for the CCD bias by using the mean bias frame obtained in the afternoon. Second, we performed the flat-fielding process. Third, we obtained the wavelength solution using the arc images taken in the afternoon. We combined multiple exposures for each target, and determined the flux errors by combining the readnoise of the CCD detector, and the Poisson noises of the target and background, using APALL task in IRAF.
Lastly, flux calibration is performed by using two standard stars, Hz~44 and HD~192281, which were observed at a similar airmass compared to that of science targets during the observing run. From the flux-calibrated 2-D spectral images, we extracted 1-D spectra along the slit. We used $\sim$1$\arcsec$ aperture size to extract spectra in order to investigate the kinematics of gas and stars as a function of radius. Using these spatially resolved 1-D spectra, we studied the connection between outflows and the large-scale radio jets. In addition, we extracted single-aperture spectra using various aperture diameters from $\sim$1$\arcsec$ to $\sim$6$\arcsec$, to study the aperture effect.

\section{Analysis}\label{analy}

%\subsection{\OIII\ velocity and velocity dispersion}\label{fit}

Using the extracted spectra, we constrain the kinematics of ionized gas using the \OIII\ line.
Following the method we adopted for our previous works based on a large sample of SDSS type 2 AGNs \citep{Woo+16}, we applied the same analysis using the spatially resolved data.  For individual spectra, we first fit the stellar absorption lines, using the penalized pixel-fitting (pPXF) code \citep{Cappellari&Emsellem04}. We used 47 MILES simple stellar population models with solar metallicity and various ages ranging from 60 Myr to 12.6 Gyr \citep{Sanchez+06}. In this process, we measured the stellar velocity and velocity dispersion for each spectrum. We adopted the stellar velocity
measured using the central spectrum as the systemic velocity of the host galaxy, which is then used to calculate the relative velocity shift of gas and stars at each radius. To demonstrate how we subtract the stellar continuum, we present the SDSS spectra in Figure 4, which cover a large spectral range with numerous stellar lines. For our MMT data, we only used one specific order, focusing on the spectral range, 5100-5400\AA\ for stellar kinematic measurements.

From the pure emission line spectra, we fitted the \OIII\ emission lines with a single or double Gaussian function by using the least-square fitting routine of Python. We applied a double-Gaussian function for the \OIII\ emission line profile if there is a significant wing or secondary component. However, if the peak of the wing or secondary component is lower than three times of the noise level in the continuum (i.e., the amplitude peak-to-noise ratio (A/N) $<$ 3), we discarded the second component as noise, and used a single Gaussian function instead. Note that this criterion was applied to all SDSS AGNs in \citet{Woo+16} to avoid any artificial fit (see Figure 4).
The signal-to-noise (S/N) of the \OIII\ emission line measured from the center of the galaxy with 1\arcsec\ aperture ranges from $\sim$20 to 40. J093552$+$612112 (hereafter J0935), J1303, and J134035$+$444817 (hereafter J1340), S/N is $\sim$20 while for J1040 and J134135$+$534444 (hereafter J1341) S/N is $\sim$30. The S/N of J120733$+$335240 (hereafter J1207) $\sim$40. Note that at the outer part of the host galaxy, the S/N becomes much lower.
%The lower limit S/N of the \OIII\ emission lines in which we apply the Gaussian fitting is S/N $>$ 2.

Based on the best-fit model of \OIII, we calculated the first moment (velocity) as,
\begin{equation}\label{eq:firstmom}
\lambda _{0} = \frac{\int \lambda f_{\lambda }d\lambda }{\int f_{\lambda } d\lambda }.
\end{equation}
Here, $f_{\lambda}$ is the flux at each wavelength. By subtracting the calculated first moment of \OIII\ from the measured systemic velocity based on stellar absorption lines, we derived the velocity shift of \OIII\ emission line.
We also calculated the second moment ($\sigma$, velocity dispersion) as,
\begin{equation}\label{eq:secmom}
\sigma_{\OIII} ^{2}=\frac{\int \lambda^{2} f_{\lambda }d\lambda }{\int f_{\lambda } d\lambda }-\lambda _{0}^{2}
\end{equation}
We corrected the measured velocity dispersion of $\sigma_{\OIII}$ for the instrument resolution of the Red Channel Echellette, $\sim$ 90 km s$^{-1}$. Figure \ref{fig:emission} shows examples of the \OIII\ emission line fits at the central pixels $\sim$3$\arcsec$ of each target.

To estimate the uncertainty of the first and second moments of each object, we created 100 mock spectra, for which the flux at each wavelength is randomized using the flux error. Then, we fitted the emission line in each mock spectrum, and adopted the 1$\sigma$ dispersion of the measured distributions as the uncertainty of the \OIII\ velocity and velocity dispersion.

%%%%%%%%%%%%%%
\begin{figure*}
\figurenum{3}
    \centering
    \includegraphics[width=0.8\textwidth]{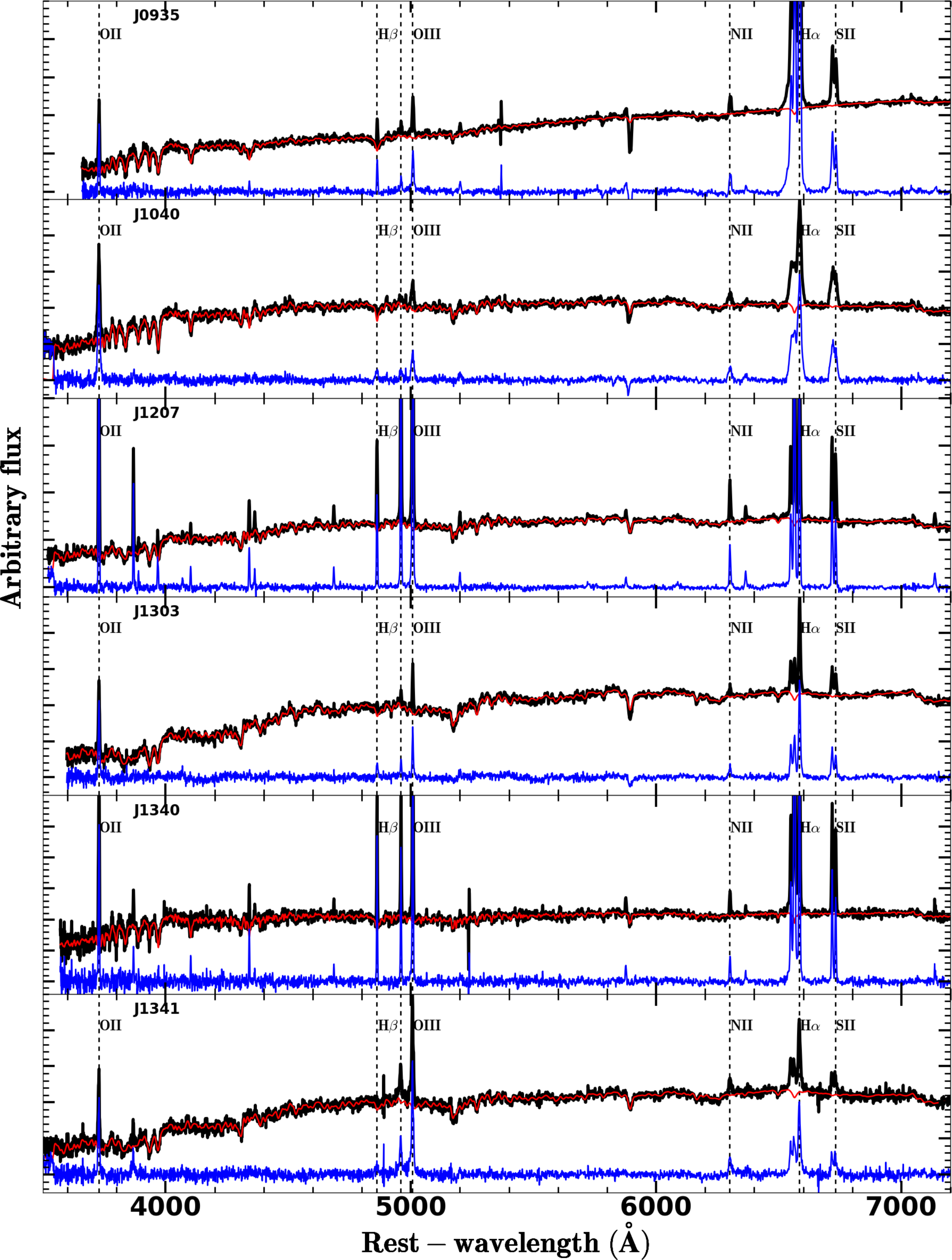}
    \caption{
    The SDSS spectra (black) of the sample, and the best-fit stellar population model (red), which provides the velocity and velocity dispersion
    of stellar component.  The residual emission line spectra (blue) show strong emission lines, including \OIII\ and \Ha.
    The dashed line indicates the location of each emission line based on the systemic velocity measured from stellar absorption lines.
    \label{fig:spec}}
\end{figure*}
%%%%%%%%%%%%%%

%%%%%%%%%%%%%%
\begin{figure*}
\figurenum{4}
    \centering
    \includegraphics[width=1.0\textwidth]{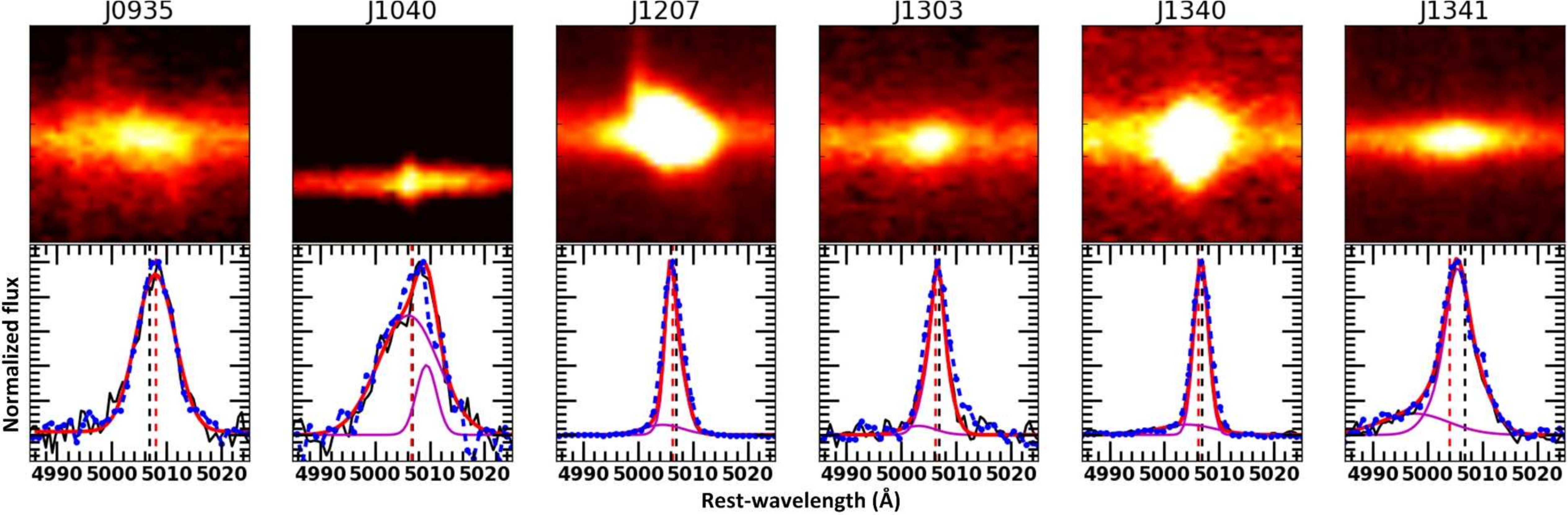}
    \caption{
   2-D spectral images around the \OIII\ line from the MMT observations, presenting a 10 \arcsec region in y-axis (top). Examples of the fitting of \OIII\ from the central pixels ($\sim$3$\arcsec$) of starlight-subtracted spectra of each target (bottom). The red-lines show the best fit while individual Gaussian components are denoted with magenta lines, and the blue-dotted lines represent the spectra from the SDSS. The black-dash vertical lines indicate the expected center of \OIII\ based on the systemic velocity, while the red-dash lines present the measured velocity of \OIII\ lines. 
    %In the case of J1040, the \OIII\ emission line is contaminated with residual OH-emission sky line.
    \label{fig:emission}}
\end{figure*}
%%%%%%%%%%%%%%

\section{Results}\label{result}

At the redshift of our targets (z $<$ $\sim$0.1), we can probe the kinematics of \OIII\ over a large scale of $\sim$10$-$20 kpc along the jet direction, with a spatial resolution of $\sim$0.8$-$2.5 kpc (i.e., 1$\arcsec$), depending on the distance to each target. We present the analysis in three parts. First, we study the kinematic signatures of \OIII\ line as a function of aperture diameter, which ranges from $\sim$1$\arcsec$ to $\sim$6$\arcsec$ at the center of the galaxy. By this step, we are able to investigate the outflow structure of the targets and the aperture effect. Second, we compare AGN outflow kinematics with accretion and radio luminosities. Third, we explore the \OIII\ kinematics as a function of a distance from the center, using $\sim$1$\arcsec$ bins. Based on these results, we will discuss the spatial connection between gas outflows and the radio emission along the jet.

\subsection{\OIII\ Kinematics as a function of Aperture Size} \label{apsize}

\begin{figure*}
\figurenum{5}
    \centering
    \includegraphics[width=0.9\textwidth]{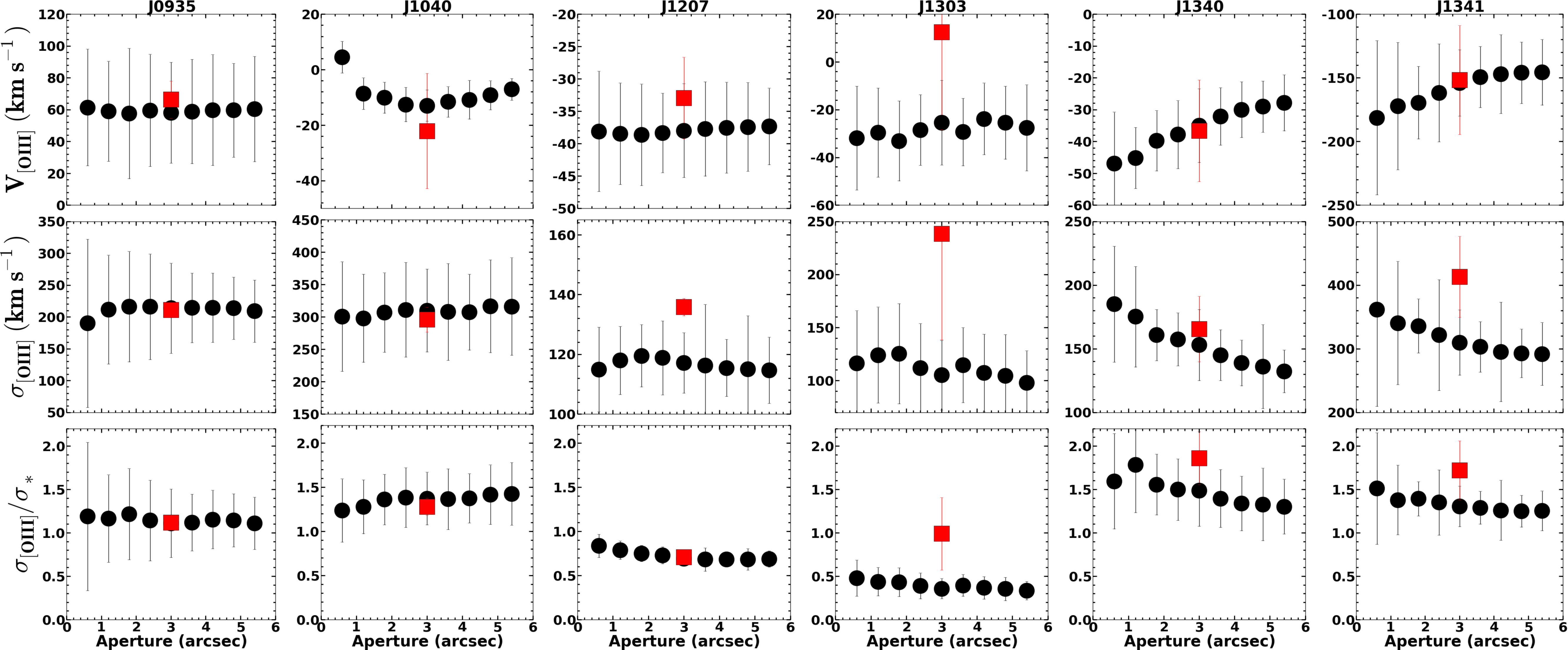}
    \caption{\OIII\ kinematics (black-dots) as a function of aperture diameter of $\sim$1$\arcsec$ to $\sim$6$\arcsec$
    from the center of all targets. The red-squares show measured values from the SDSS spectra (3$\arcsec$). The vertical lines indicate the measurement errors.
    \label{fig:apsize}}
\end{figure*}

In Figure \ref{fig:apsize}, we present the \OIII\ velocity shift with respect to the systemic velocity, and velocity dispersion of each target as a function of aperture diameter. We used various aperture sizes, ranging from $\sim$1$\arcsec$ to $\sim$6$\arcsec$ to investigate the kinematic difference between inner and outer part of each galaxy, and to compare with the kinematic measurements based on the SDSS spectra, which are obtained through
3\arcsec\ fiber aperture size.  We see five AGNs show a blueshifted \OIII\ while one AGN, J0935, shows a redshifted \OIII, with a mean value of $\sim$60 km s$^{-1}$. Since the measured velocity shift is based on the flux-weighted spectrum, the velocity shift would be zero if the outflows have a biconical shape, and the motions of the approaching and receding gas are canceled each other. However, since the dusty stellar disk preferentially hides a part of the cone, we observe a velocity-shifted \OIII\ line as demonstrated by \citet{Bae&Woo16}, who simulated the emission line profile and velocity shift based on the 3-D outflow models combined with a dusty stellar disk.

As a function of aperture diameter, we find that four targets, namely, J0935, J1040, J1207, and J1303 show a relatively constant \OIII\ velocity shift,
indicating that there is no significant radial change of the ionized gas kinematics, and the central part of the narrow line region is dominating
in terms of flux.  
 It is also possible that if the ionized gas follows the rotation due to the host galaxy's gravitational potential (see the examples of J0935 and J1207 in Figures 4 and 7), then the symmetric velocity distribution (i.e., negative and positive velocities) would not contribute to the velocity shift of the total line profile in the flux-weighted aperture spectrum. 

As we described in Section 2.1, the gas to stellar velocity dispersion ratios of these objects suggest that
ionized gas outflows are not strong and mainly governed by the gravitational potential of host galaxies. This is supported by the insignificant
change of the \OIII\ velocity shift with an increasing aperture size. In contrast, J1340 and J1341, which have strong outflows indicated by a large gas to stellar velocity dispersion ratios (i.e., $\sigma_{OIII}$ / $\sigma_*$ $\sim$2), present a clear decreasing trend, suggesting that outflows are strong at the center and slowing down outwards, which is also manifested in Figure 7. 

Turn to the velocity dispersion of \OIII, we see the similar trends as the case of velocity shift. Four targets (J0935, J1040, J1207, and J1303) with weak or no outflows show a constant value as a function of the aperture size, while two targets (J1340 and J1341) with strong outflows show a decreasing trend of \OIII\ velocity dispersion with an increasing aperture diameter (see also \S~4.3). This steep decrease of \OIII\ velocity dispersion is presumably
due to the fact that gas outflows are strong in the center and become weaker radially. Note that we see the similar steep decrease of \OIII\ velocity dispersion as a function of radial distance in our IFU studies on low-z type 2 AGNs \citep{Karouzos+16, Karouzos+16b, Bae+17}.  

We also compare the velocity shifts and velocity dispersions of \OIII\ which measured from the SDSS spectra with that of our results.
We find a consistency between the measurements based on our MMT data and those based on the SDSS spectra, except for J1303, which shows an opposite sign of the \OIII\ velocity shift.
In the case of velocity dispersion, the measurement based on SDSS spectrum is much larger than that based on the MMT spectrum.
This is already hinted by the difference in the line profiles as the \OIII\ line in the SDSS spectrum is broader than that in the MMT spectra
although the different spectral resolutions also play a role in the apparent discrepancy (see Figure \ref{fig:emission}).
However, since the error on the SDSS measurements, which are consistently measured based on Monte Carlo simulations, are very large due to the low S/N, and the two measurements are consistent within the error, we did not further investigate the difference.

%%%%%%%%%%%%%%
%%%%%%%%%%%%

\begin{figure}
\figurenum{6}
    \centering
    \includegraphics[width=0.45\textwidth]{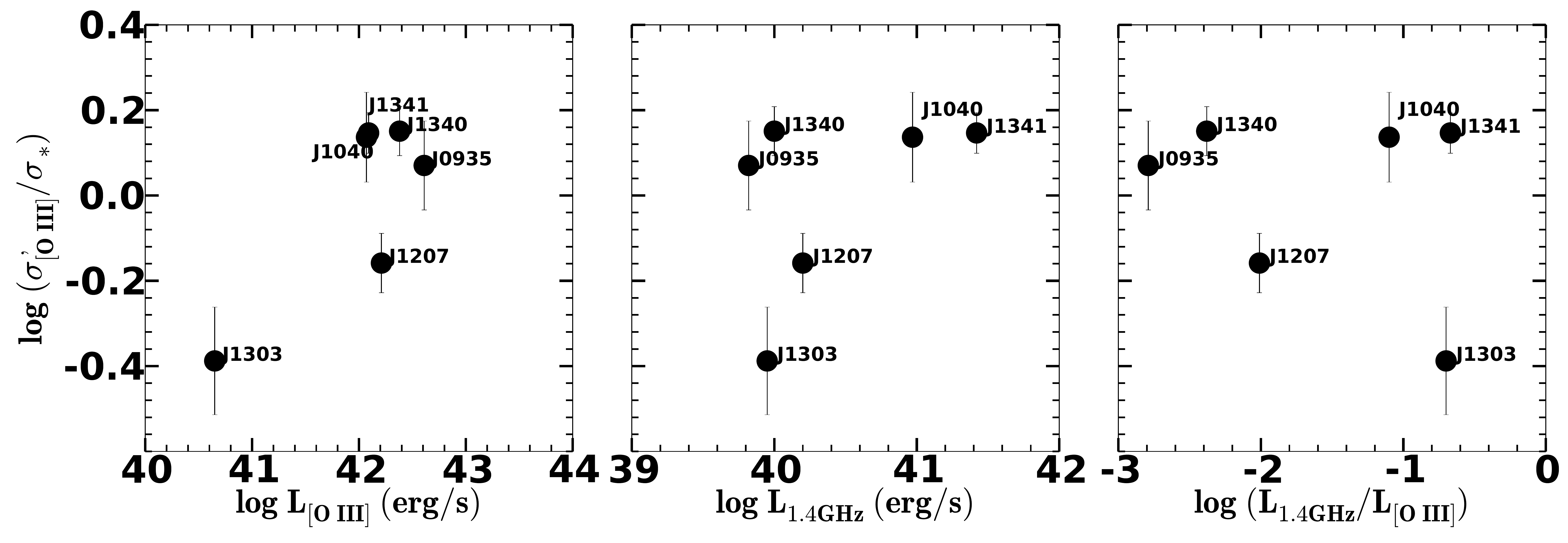}
    \includegraphics[width=0.45\textwidth]{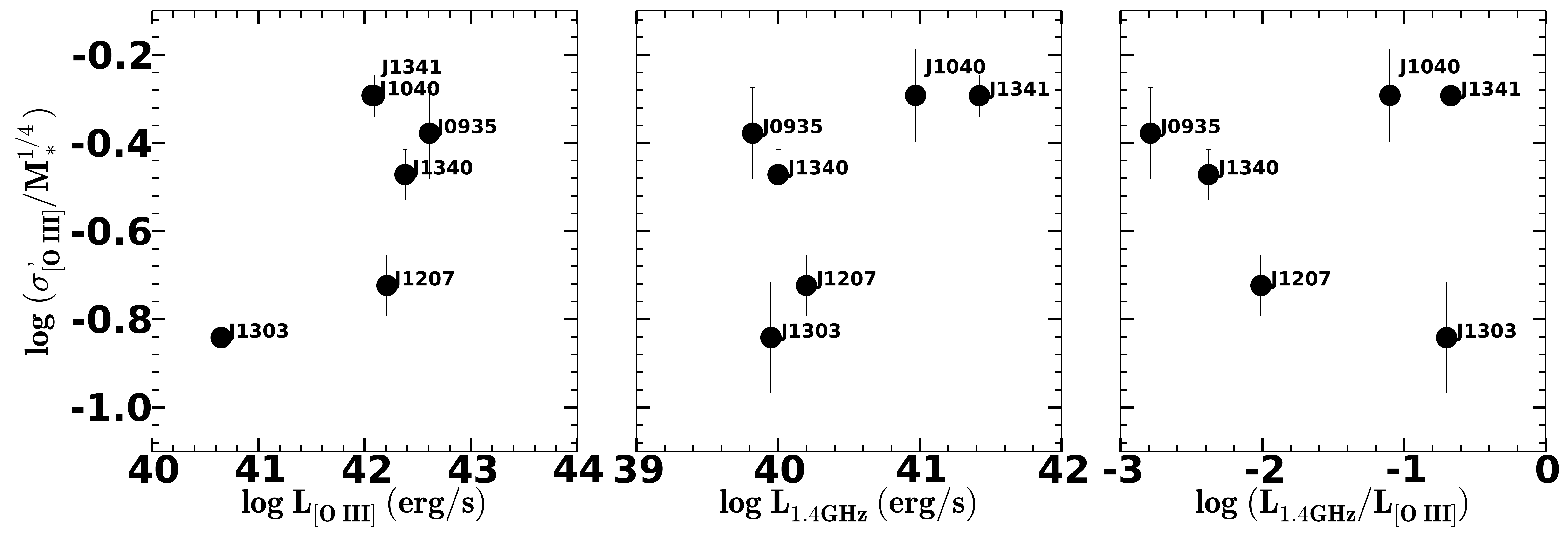}
    \caption{Comparison of non-gravitational \OIII\ velocity dispersion (3$\arcsec$) with the \OIII\ luminosity (extinction correction), the radio luminosity $\mathrm{L_{1.4GHz}}$, and radio-to-\OIII\ luminosity ratio. In top panel,
    $\mathrm{\sigma^{'}_{[OIII]}}$ is normalized for stellar velocity dispersion $\mathrm{\sigma_{*}}$.
    In bottom panel, $\mathrm{\sigma^{'}_{[OIII]}}$ is normalized for the stellar mass to the 1/4 power, $\mathrm{M_{*}^{1/4}}$.
    \label{fig:lum}}
\end{figure}

%%%%%%%%%%%%%%%%%

\begin{figure}
\figurenum{7}
    \centering
    \includegraphics[width=0.23\textwidth]{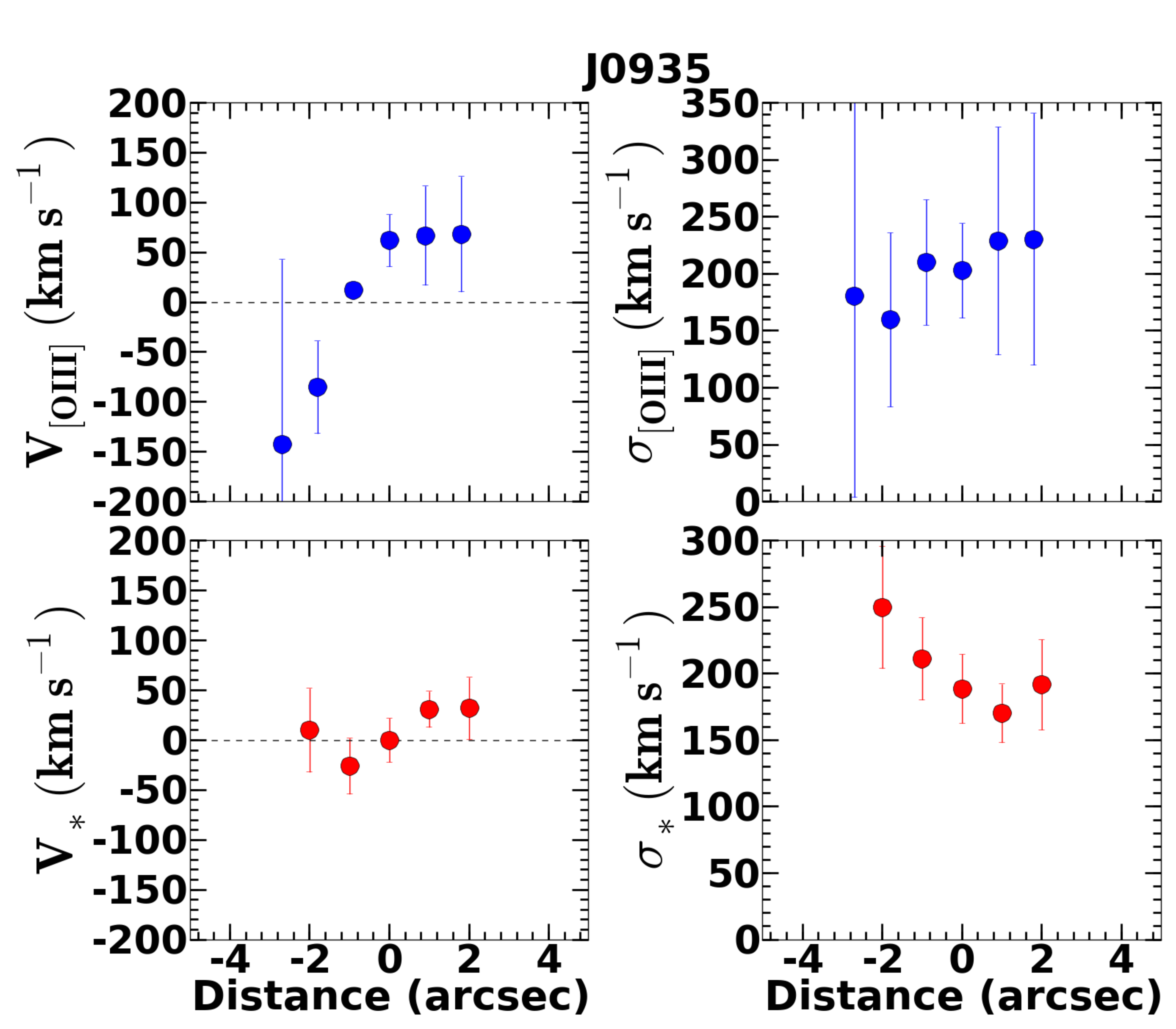}
    \includegraphics[width=0.23\textwidth]{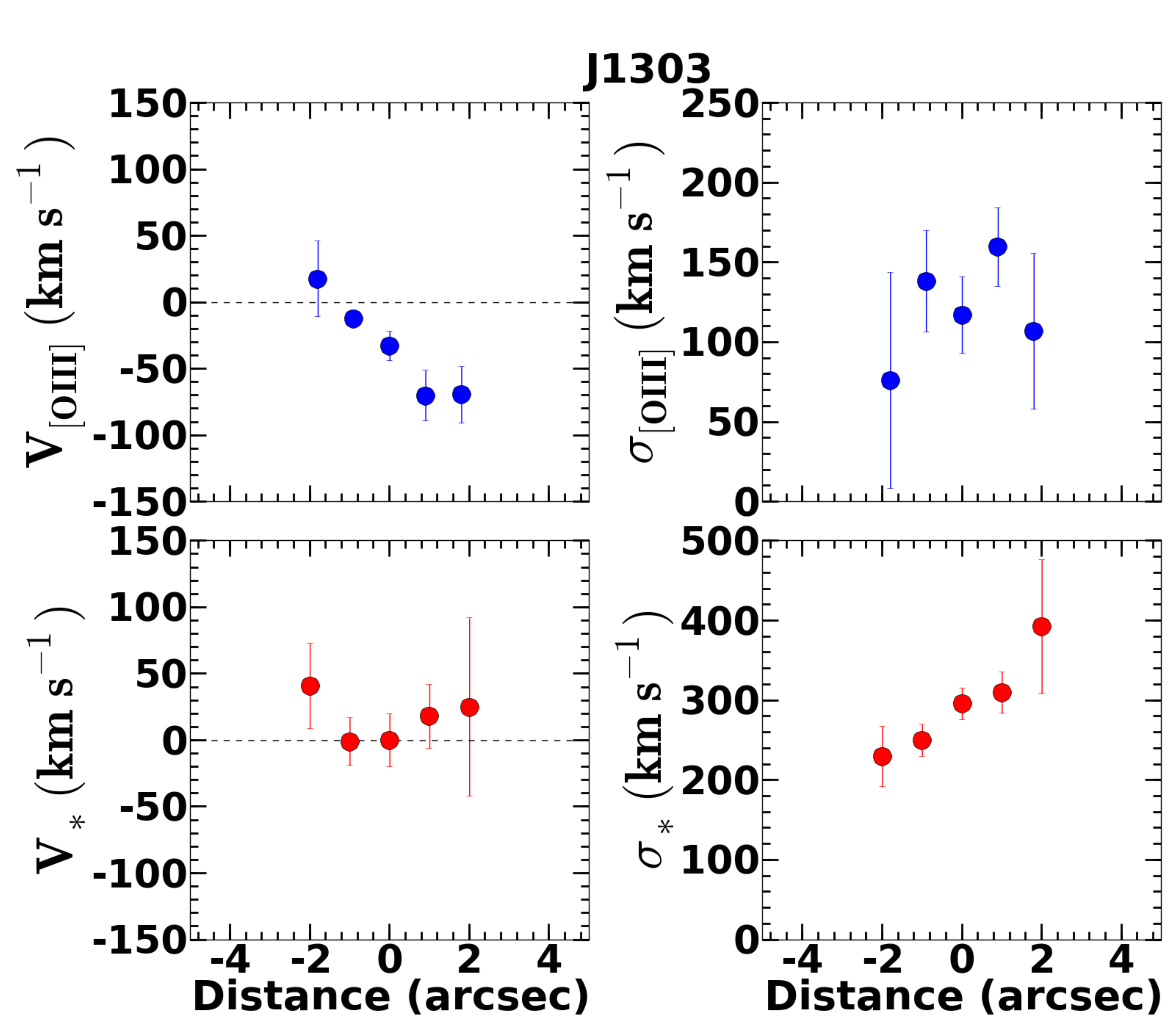}\\
    \includegraphics[width=0.23\textwidth]{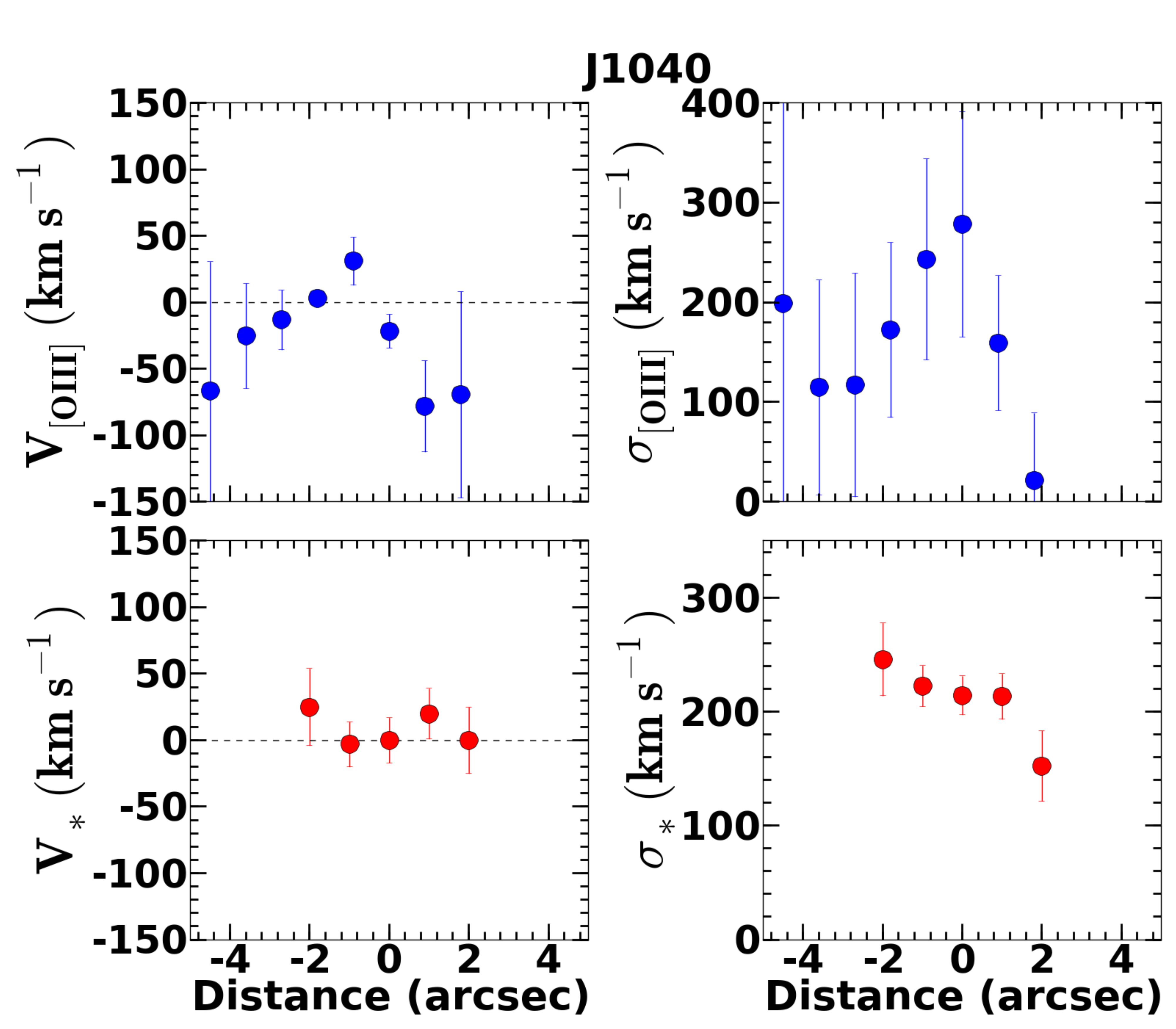}
        \includegraphics[width=0.23\textwidth]{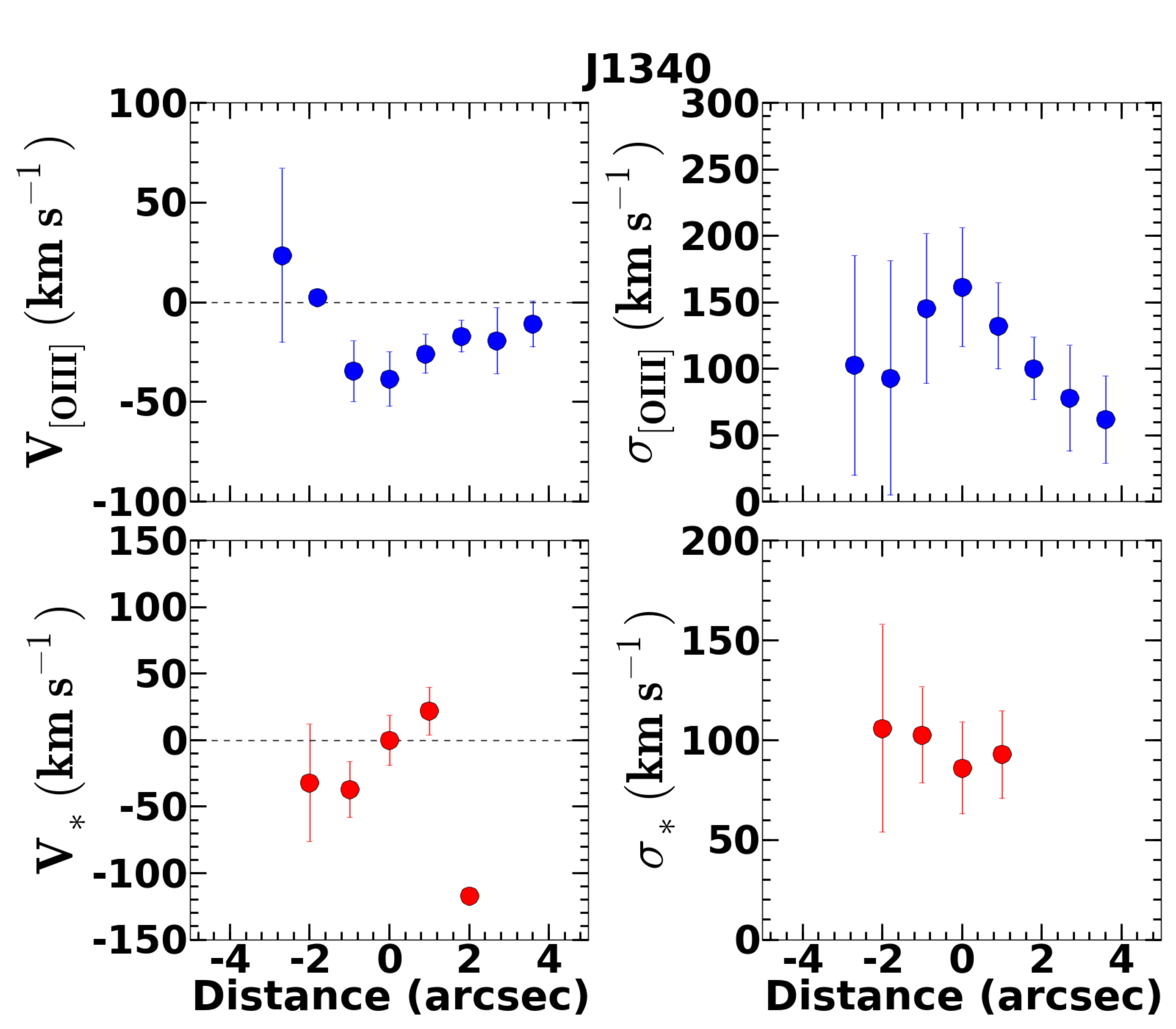}\\
     \includegraphics[width=0.23\textwidth]{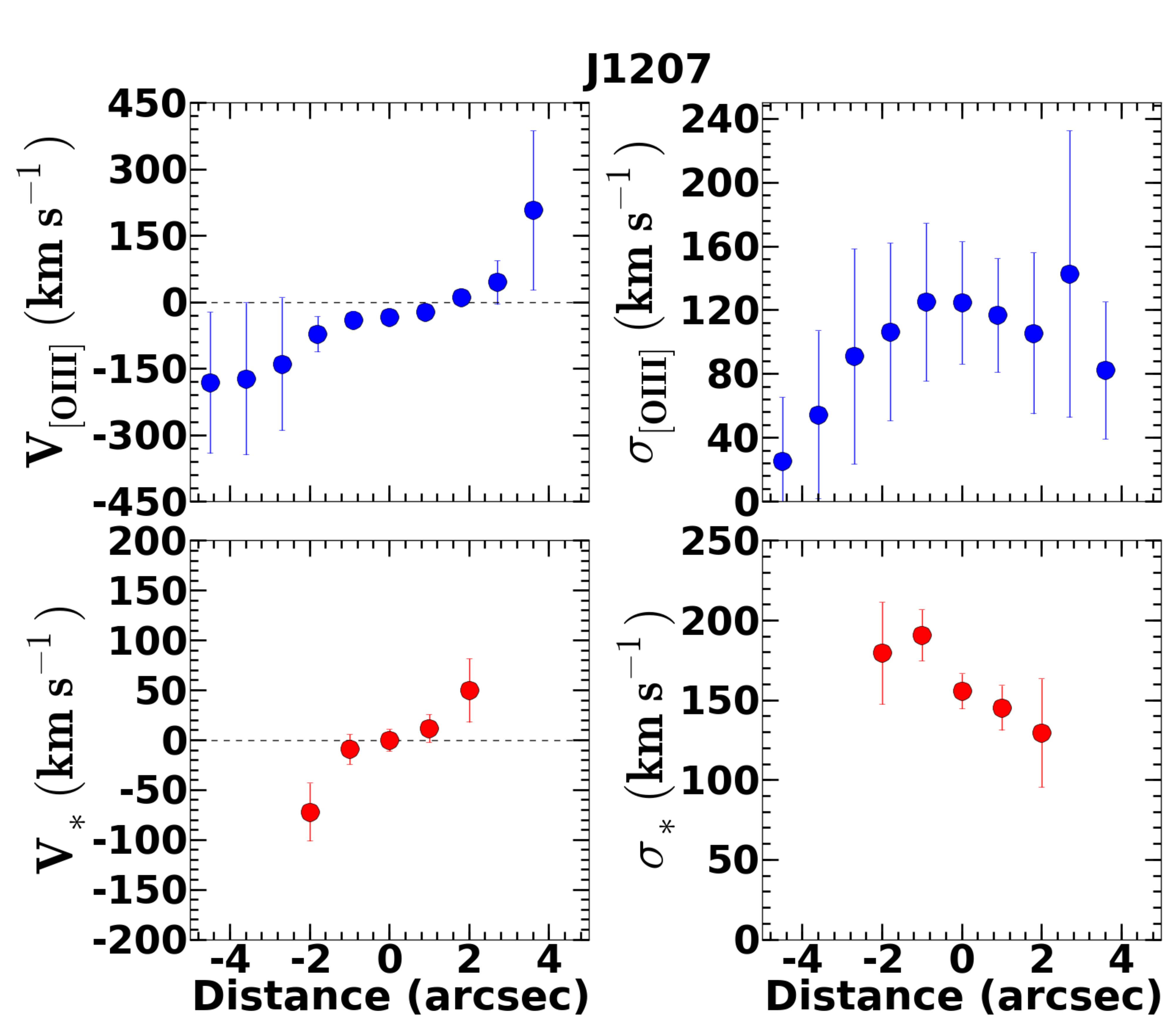}
         \includegraphics[width=0.23\textwidth]{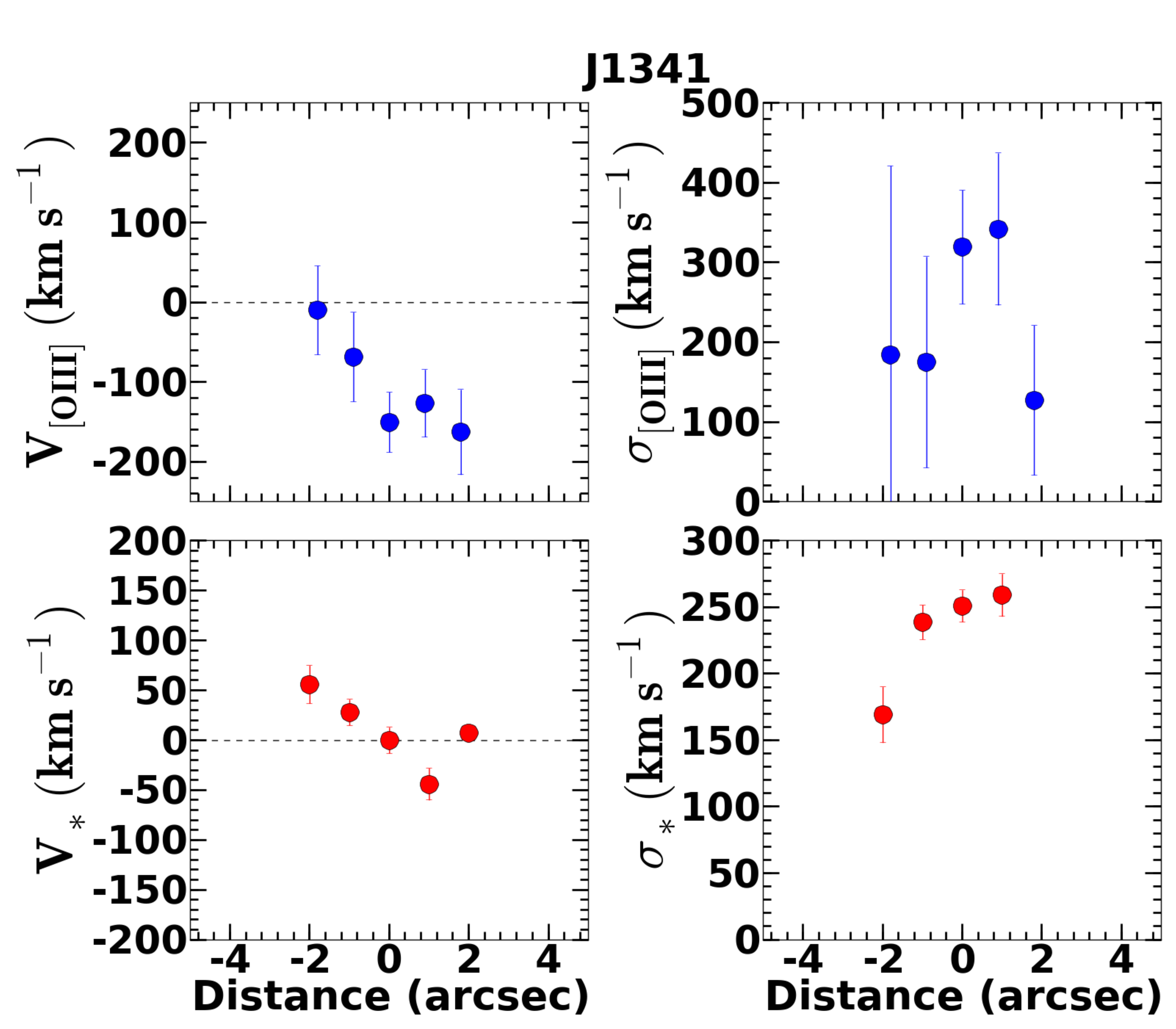}
    \caption{\OIII\ kinematics (blue-dots) as a function of spatial resolution distributions. The bottom plots present the measurements of stellar velocity
    and velocity dispersion (red-dots). %The purple-squares show the measured values from the SDSS spectra (3$\arcsec$).
    \label{fig:spa1}}
\end{figure}
%%%%%%%%%%%%%%%%%%%

\subsection{AGN outflow vs. luminosity} \label{lum}

In this section, we compare the non-gravitational kinematic component of \OIII\  with the extinction-uncorrected luminosity of \OIII\ and the radio luminosity $\mathrm{L_{1.4GHz}}$. We use the sum of the velocity and velocity dispersion in quadrature, $\mathrm{\sigma^{'}_{[OIII]}}$, which can be interpreted as the velocity dispersion corrected for dust extinction and bicone inclination effects \citep[for details see][]{Bae&Woo16}. Then, the $\mathrm{\sigma^{'}_{[OIII]}}$ is normalized by the stellar velocity dispersion $\mathrm{\sigma_{*}}$ or the stellar mass to the 1/4 power $\mathrm{M_{*}^{1/4}}$ to indicate the relative strength of the non-gravitational kinematic component.

Figure \ref{fig:lum} displays the comparisons between the $\mathrm{\sigma^{'}_{[OIII]}}$ velocity normalized by the stellar velocity dispersion (top panels) and stellar mass (lower panels) with \OIII\ luminosity, the radio luminosity $\mathrm{L_{1.4GHz}}$, and radio-to-\OIII\ luminosity ratio. In the case of the $\mathrm{\sigma^{'}_{[OIII]}}$ normalized by the stellar velocity dispersion, we see that two weak-outflow targets (J1207 and J1303) with low $\log$ $\mathrm{\sigma^{'}_{[OIII]}}$/$\mathrm{\sigma_{*}}$  are located somewhat separately at lower part of the plot, compared to other targets with intermediate or strong outflow signatures (i.e., $\log$ $\mathrm{\sigma^{'}_{[OIII]}}$/$\mathrm{\sigma_{*}}$ $\sim$ 0.2 (J0935, J1040, J1340, and J1341). Although the sample is small, we see a significant trend between the non-gravitational kinematic components and the \OIII\ luminosity, indicating that the outflow kinematics are connected with the accretion power as manifested by \OIII\ luminosity. The trend between \OIII\ luminosity and outflow velocity is consistent with those found in a statistical study of type 2 AGNs by \cite{Woo+16}. In contrast, for the comparison with the radio luminosity $\mathrm{L_{1.4GHz}}$, and ratio of radio-to-\OIII\ luminosity, we find no significant trend, suggesting that the non-gravitational kinematic component of \OIII\ is not directly connected to the radio activities. Similarly, we found the consistent results in comparing the
$\mathrm{\sigma^{'}_{[OIII]}}$ normalized by stellar mass and the \OIII\ luminosity, the radio luminosity, and the radio-loudness.

%In addition, in case of the outflow kinematics normalized by stellar mass, we see similar results as in the case of normalization by stellar velocity dispersion. Two targets with weak-outflows has $\log$ $\mathrm{\sigma^{'}_{[OIII]}}$/$\mathrm{\sigma_{*}}$ $<$ -0.7. And, Four targets which have intermediate and strong outflows, $\log$ $\mathrm{\sigma^{'}_{[OIII]}}$/$\mathrm{\sigma_{*}}$ $\sim$ -0.3. Similarly, we also found no significant trends for those comparisons with the radio lunimosity.
 %%%%%%%%%%%%%%%%%%%

\subsection{Spatially Resolved \OIII\ Kinematics of individual objects} \label{apspa}

In this section, we present the spatially resolved stellar and \OIII\ kinematics along the jet direction using a spatial bin size of $\sim$1$\arcsec$. Figure \ref{fig:spa1} shows the velocity with respect to systemic velocity  and velocity dispersion, respectively for \OIII\ (top) and stars (bottom).

First, we discuss the two objects with no outflow signatures in \OIII, namely, J1207 and J1303. J1207 is an early-type galaxy with a clear sign of rotation at the central part, where the S/N of stellar absorption line is enough to measure stellar velocities. \OIII\ shows a similar trend in the velocity profile, indicating that gas and stars are mainly governed by the gravitational potential of the host galaxy as expected. In the case of the velocity dispersion profile, \OIII\ shows a radially decreasing trend, however, the amplitude of velocity dispersion is somewhat lower than that of stellar velocity dispersion, suggesting that no additional kinematic component is present in the jet direction. J1303, which is classified as the brightest cluster galaxy in AbelL 1168 \citep{Ascaso+11}, shows very high stellar velocity dispersions over 300 \kms\ at the center without a clear rotation feature.
In contrast, \OIII\ lines are relatively weak and velocity and velocity dispersion of \OIII\ do not show any gas outflow signature.
As expected, these AGNs with no significant outflow signature suggest that radio activity does not trigger or enhance gas outflows.

%we see that the trend of the \OIII\ kinematics of the two targets are similar. The velocity shifts display the rotation curves. The patterns indicate the rotations of gas motions. There are blueshifted and redshifted emission gas (relative to the systemic velocity). The trends of stellar velocity dispersion display the maximum values, $\sim$ 130 km s$^{-1}$ at the center of the galaxy, and show lower values, $\sim$ 60 km s$^{-1}$ at distance from $\pm$2 to $\pm$4$\arcsec$. In case of the stellar kinematics, the radial stellar velocities display rotation curves, indicating the presence of the bulge component. The rotational trend of the stellar velocity of J1207 show similar behavior with that of the \OIII\ velocity shifts. It is hard to say about the stellar velocity trend of J1303 due to the low signal-to-noise (SN) of the data. The stellar velocity dispersions of both targets show consistent value with that of SDSS spectra. The trends of stellar velocity dispersion show constant values due to the low signal-to-noise of the data.

Second, we investigate the two weak outflow AGNs, namely, J0935 and J1040. The morphology of these host galaxies are disturbed and no clear rotation is detected in the stellar kinematics. J0935 is classified as ultra luminous infra-red galaxy (ULIRG) and the spectrum shows a strong dust extinction in short wavelengths (see Figure 3). Thus, we also expect strong star formation activities, which may also trigger gas outflows.
In the case of velocity dispersion, stars and gas show very different radial trend, indicating that gas and stars are kinematically decoupled as expected if the host galaxy is in a merging process. J1040 (4C 30.19) also shows a slightly disturbed morphology, but the color of the host galaxy indicates that it is an early-type galaxy without a clear rotation in stellar kinematics. \OIII\ shows a larger velocity dispersion than that of stars at
the center and radially decrease. This trend may be interpreted as decelerating outflows although it is equally viable that gas follows
the potential of the host galaxy.

Third, we turn to the two strong outflow AGNs, J1340 and J1341, for which we detect large velocity dispersions of \OIII\ compared to stars, particularly at the center. In the case of J1340, which is a flat-spectrum radio galaxy, we find a disturbed morphology and indication of star formation based on galaxy color in the SDSS image.  The velocity profile indicates that \OIII\ lines are mostly blueshifted, suggesting gas outflows toward the line-of-sight while a weak rotation is present in stellar velocity profiles.  \OIII\ velocity dispersion is much higher than that of stars,  particularly at the center, indicating the presence of additional kinematic component, i.e., outflows. J1341 shows similar trends that \OIII\ velocity dispersion is larger than stellar velocity dispersion. Also, \OIII\ lines show blueshift, indicating outflows, while stellar kinematics show a clear rotation.

Among these 6 AGNs, two radio galaxies, J1040 and J1341 have the largest radio luminosity. If there is a connection between radio activity and ionized gas outflows, we may see a strong gas outflows in these two objects. However, we do not find a significant difference between strong and weak radio AGNs in terms of the \OIII\ kinematics, suggesting that ionized gas outflows are mainly due to the AGN disk radiation rather than large-scale jet activity.

\subsection{AGN-Photoionisation size} \label{outflow}

In this section, we investigate the size of the NLR for our sample AGNs and compare it with \OIII\ luminosity.
The photoionization size ($\mathrm{R_{NLR}}$) can be probed based on the spatial distribution of ionized gas, which is manifested by the surface brightness distribution in the narrow band images with the \OIII\ filter or by the spatial distribution of \OIII\ emission in long-slit or integral field
spectroscopy \citep[e.g.][]{Bennert+06a, Bennert+06b, Husemann+13, Oh+13, Karouzos+16, Bae+17}. We check the \OIII/\Hb\ ratio as a function of radius, confirming that the ionizing source is the central AGN in each object. An exception is J0935, which is an Ultra Luminous Infra-Red galaxy. One side of this galaxy to north-east direction shows a much weaker \OIII/\Hb\ ratio (i.e., $<$ 3), indicating that the ionizing source is star formation. Nevertheless, since the south-west direction shows the \OIII/\Hb\ ratio $\geq$ 3, the photoionization size measured in 2-D distribution is still acceptable. 

Using the 2-D spectral images, we fitted the spatial distribution of the flux in the spectral range of \OIII\ with a Gaussian model, to measure the size of the NLR. In comparison,  we also measured the FWHM of the flux distribution using the 2-D spectral images of flux calibration stars, to determine the seeing size. The seeing during our observations was $\sim$1$\arcsec$, which was good enough to resolve the \OIII\ region since the FWHM of the observed \OIII\ flux distribution ranges from $\sim$1.5 to 2.9$\arcsec$ for the sample. In addition, we checked the spectra extracted from each radius and confirmed that the \OIII\ emission is spatially resolved.

The spatially resolved \OIII\ emission region is detected out to a radial distance of $\sim$2$-$4$\arcsec$. For the most distant target J1341 in the sample, the kinematics of \OIII\ can be still measured out to a distance of 2$\arcsec$, which corresponds to the linear scale of $\sim$5 kpc. In a number of previous studies of the \OIII\ photoionization size in the literature, \OIII\ NLR size is measured from the Gaussian component using the FWHM of \OIII\ distribution (\citealp{Bennert+06a}; \citealp{Bennert+06b}; \citealp{Rodriguez+13}; \citealp{Karouzos+16}).
We defined the radius of the $\mathrm{R_{NLR}}$ by dividing the FWHM$_{\OIII}$ by a factor of two. Considering the spatial resolution, we then subtracted the seeing size from the measured FWHM$_{\OIII}$ in quadrature
(see Table \ref{tab:outflow}).  The measured $\mathrm{R_{NLR}}$ size is in the range of 0.9$-$1.6 kpc. To estimate the errors of the $\mathrm{R_{NLR}}$ sizes, we constructed 100 mock spatial distributions of the \OIII\ flux, by adding a randomized error to each pixel,
then we used the same method to measure the $\mathrm{R_{NLR}}$ sizes for each mock of the \OIII\ flux distributions. We adopted the 1$\sigma$ dispersion of the size distribution as the error of $\mathrm{R_{NLR}}$ size. In addition, we also add 10$\%$ of the $\mathrm{R_{NLR}}$ and a half of the pixel size ($\sim$0.15$\arcsec$) in order to take into account for the uncertainties of the seeing size and the spatial sampling.

\begin{figure}
\figurenum{8}
    \centering
    \includegraphics[width=0.48\textwidth]{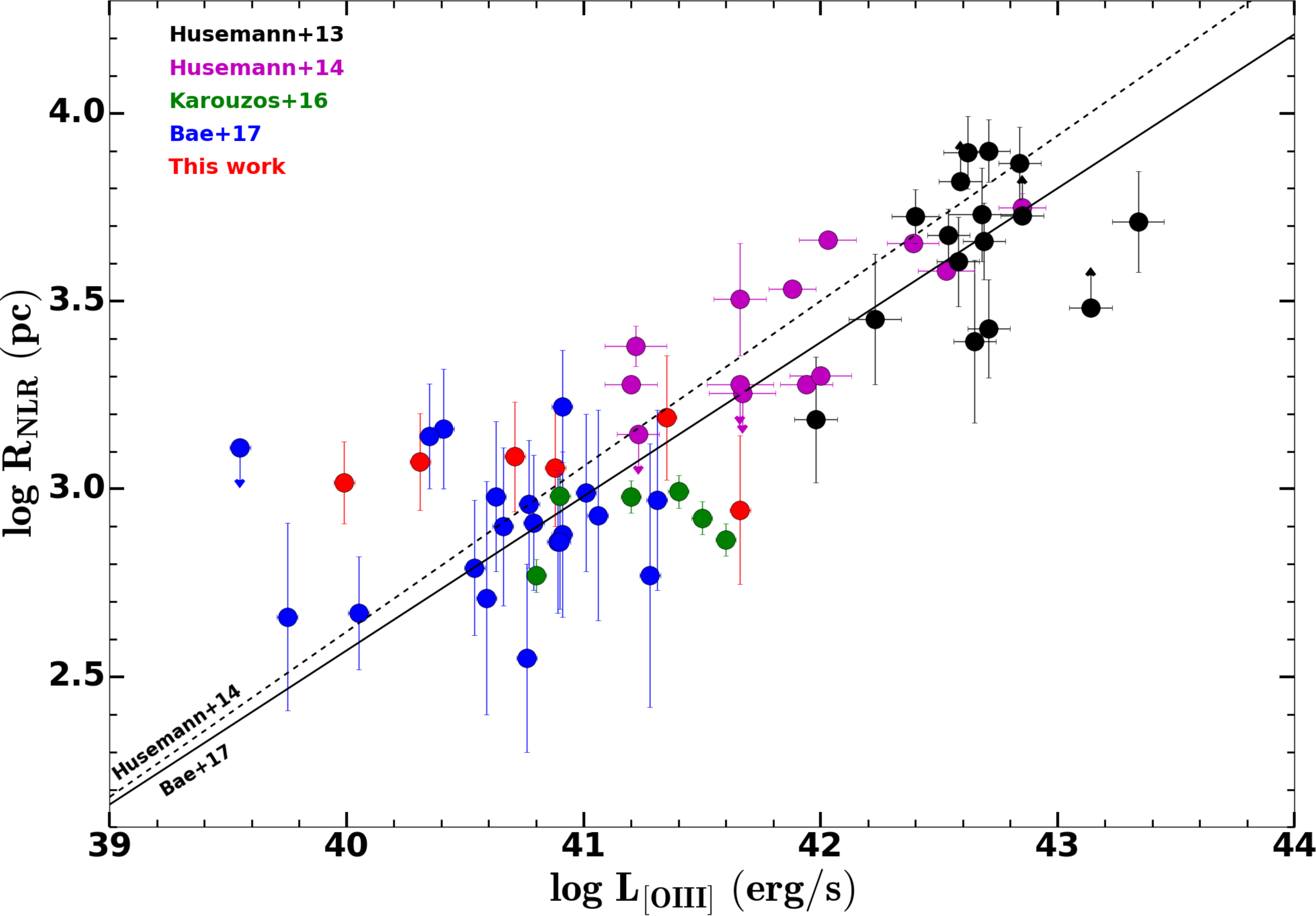}
    \caption{The photometric size-luminosity relations for AGNs. Black and magenta dots show the measured sizes from \citet{Husemann+13, Husemann+14}.
    Green-dots present the results from \citet{Karouzos+16}. The measured values from \citet{Bae+17} are denoted with blue-dots. Our measured \OIII\ sizes are shown in
    red-dots. In this plot, the \OIII\ luminosity is the extinction-uncorrected values. Dash and solid lines present the slope values from linear regression of
    \citet{Husemann+14} and \citet{Bae+17} .
    \label{fig:size}}
\end{figure}

In Figure \ref{fig:size} we compare the measured NLR size based on \OIII\ with the \OIII\ luminosity for our sample, combined with various AGNs selected from literature, for which the photoionization size was measured with a consistent manner (\citealp{Husemann+13, Husemann+14}; \citealp{Karouzos+16}; \citealp{Bae+17}). Note that since all these measures are dust-uncorrected \OIII\ luminosity, we use \OIII\ luminosity without correcting for dust obscuration. We find that our radio AGNs in general follow the same trend between the size of NLR and \OIII\ luminosity while AGNs at the low luminosity end show slightly larger scatter. However, we should point out that the jet direction may not be correctly derived since for some objects the radio image does not show a clear jet direction. Also, it is possible that the jet axis defined in the 1.4 GHz image may not be aligned with the ionized gas outflow direction if there was a time lag between the formation of the jet and the outflows. Since our analysis depends on the longslit data with a specific P.A., the estimated size of the NLR based on \OIII\ may be underestimated if the slit is off from the outflow direction. 

The flux distribution of \OIII\ is often used to define the size of the NLR. However, the extent of the photoionization region is different from that of gas outflows, since gas can be ionized by the central source while the kinematics of the ionized gas can simply follow the gravitational potential of the host galaxy without showing any outflows. Therefore, \citet{Karouzos+16} and \citet{Bae+17} measured the size of the outflows based on the \OIII\ kinematics compared to stellar kinematics, instead of measuring the extent of the photoionization region. A detailed comparison between the photoionization and outflow sizes will be presented in the future (Kang et al. 2017 in preparation). While the outflow size measured based on the spatially resolved kinematics provides much better representation of the outflow energetics, it requires high quality data with a good spatial resolution, which are missing in our MMT observations. Future IFU observations with high S/N may overcome the limitations of our analysis based on MMT longslit spectra.

%%%%%%%%%%%%%%
%%%%%%%%%%%%

%%%%%%%
\begin{table}
\begin{center}
\caption{Photoionisation size} \label{tab:outflow}
\begin{tabular}{cccc}
\tableline\tableline
ID	&	FWHM$_{\OIII}$   &   V$_{\OIII}$   &   	$\mathrm{R_{NLR}}$
\\
   &  (Km s$^{-1}$)   &  (Km s$^{-1}$)  &   (Kpc) \\
   (1)  &  (2)  &  (3)  & (4) \\
\tableline
J093552$+$612112	&	502$\pm$115	&	60$\pm$33	&	1.0$\pm$0.3 \\
J104030$+$295758	&	746$\pm$170	&	-9$\pm$5	&	1.2$\pm$0.4	\\
J120733$+$335240	&	270$\pm$42	&	-37$\pm$7	&	0.9$\pm$0.4	\\
J130347$+$191617	&	247$\pm$92	&	-25$\pm$15	&	1.2$\pm$0.4	\\
J134035$+$444817	&	320$\pm$78	&	-29$\pm$8	&	1.1$\pm$0.4	\\
J134135$+$534444	&	690$\pm$89	&    -146$\pm$24	&	1.6$\pm$0.6	\\
\tableline
\end{tabular}
\tablecomments{(1) name; (2) FWHM of \OIII\ ; (3) velocity shifts with respect to the systemic velocity; (4) estimated radius of the \OIII\ region. }
\end{center}
\end{table}
%%%%%%%%%%%

\section{Discussion and Conclusions}\label{sum}

Various studies in the literature showed the connection between gas kinematics and radio emissions. For example, \citet{Mullaney+13} reported that in AGNs with moderate radio-luminosity, the radio-luminosity $\mathrm{L_{1.4GHz}}$ has strong effects on the \OIII\ profile. The morphology and position angle of the radio emission and the ionized gas emission have strong connections. In the case of the targets in which the jet and ionized gas emissions are aligned, the radio jets may be responsible for driving outflows and shock ionization, hence driving the correlations between radio emissions and gas kinematics (e.g., \citealp{Veilleux91}; \citealp{Bower+95}; \citealp{Zakamska&Greene14}).
In a radio-loud AGN 4C12.50  fast outflows in HI gas are detected with an extremely blue-shifted velocity, $\sim$1000 km s$^{-1}$ relative to the systematic velocity at the location of the southern jet. These outflows are considered as a result of the interaction between radio jet and the ISM \citep{Morganti+13}. By using near-infrared data, \citet{Tadhunter+14} reported the highly distorted kinematic profiles in the H$_{2}$ 1-0 S(1) $\lambda$2.128$\mu$m line, showing a clear evidence that the interaction between expending radio jet and the ISM is the reason of accelerating the molecular gas in the western radio lobe of the Seyfert galaxy IC~5603.

We expect that if large-scale radio jets are responsible for triggering or enhancing gas outflows, there may be significant kinematic signatures along the jet direction or connection with radio luminosity. However, from the results in Section 4.2 and 4.3, we find no strong evidence that \OIII\ kinematics are connected with radio activity along the jet direction for our small sample of low redshift radio sources.
We note that there could be a strong resolution effect. The spatial resolution of our optical spectroscopy with the MMT is about 1\arcsec\ while
the jet morphology was probed based on 5 \arcsec\ spatial resolution with the FIRST. As we determined the PA of the slit to explore the ionized
outflow based on the FIRST radio images, it is possible that the direction of radio jets in small scales can be different since jets are known
to bend and the direction of the jet changes at different spatial scales (e.g., \citealp{Zakamska&Greene14}).
While an integral field spectroscopy may provide better constraints on the connection between jet and gas outflows, our investigation is limited by the fixed direction of long-slit spectroscopy.

~

We present the analysis of \OIII\ kinematics of six high radio luminosity type 2 AGNs using the Red Channel Cross Dispersed Echellette Spectrograph. Our main results are as follows,

\medskip

1. By analyzing \OIII\ kinematics as a function of aperture diameter, we find that the \OIII\ kinematics show an increasing trend in velocity and decreasing trend in velocity dispersion in the case of the AGNs with strong outflow signatures. These trends indicate that outflows are strong at the center and become weaker radially. In contrast for the AGNs with no or weak outflow signatures, the \OIII\ kinematics show no trend as a function of the aperture size, suggesting that ionized gas outflows are not strong and \OIII\ kinematics are mainly governed by the gravitational potential of host galaxies.

2. By investigating \OIII\ kinematics along the radio jet direction with a spatial scale of $\sim$1$\arcsec$, we find no evidence of the connection between ionized gas outflows and the large-scale radio jet emissions in our low-redshift radio AGNs. The ionized gas kinematics mainly correlate with AGN disk radiation rather than the jet activity.

3. We find an increasing trend between outflow kinematics and \OIII\ luminosity, indicating that for high luminosity AGNs have stronger outflows,
which is consistent with the result based on the statistical study of a large sample of optical type 2 AGNs by \cite{Woo+16}. This result indicates that more luminous AGNs show stronger outflows, regardless of radio activities.

4. By measuring the spatial distribution of the \OIII\ flux, we determine the size of the AGN-photoionization, which ranges from 0.9 to 1.6 kpc. These measurements are consistent with the size-luminosity relation of optical type 2 AGNs, suggesting that radio AGNs also follow the same relation.

\acknowledgments
We thank the anonymous referee for various suggestions, which improved the clarity of the paper. 
This work was supported by the National Research Foundation of Korea grant funded by the Korea government (No.2017R1A5A1070354  and No. 2016R1A2B3011457).

\bibliographystyle{apj}

\end{document}